\documentclass[
 reprint,
 amsmath,amssymb,
 aps,nofootinbib
]{revtex4-1}
\usepackage[T1]{fontenc}
\usepackage{graphicx}
\usepackage{dcolumn}
\usepackage{bm}
\usepackage{color}
\begin{document}

\preprint{APS/123-QED}

\title{Exotic Hadrons in the $\Lambda_b \rightarrow J/\psi \ \phi \ \Lambda$ Decay}

\author{Volodymyr Magas}
\author{\`{A}ngels Ramos}
\author{Rahul Somasundaram}
\affiliation{
Departament de Fisica Quantica i Astrofisica and Institut de Ciencies del Cosmos\\
University of Barcelona, 08028-Barcelona, Spain}
\author{J\'ulia Tena Vidal}
\affiliation{Department of Physics, University of Liverpool, 
The Oliver Lodge Laboratory, Liverpool, L69 7ZE
United Kingdom}
\date{\today}

\begin{abstract}
We study the weak decay of the $\Lambda_b$ baryon into $J/\psi\ \phi\ \Lambda$, a process that is particularly well suited to analyze the physics of some of the recently observed or theoretically predicted exotic hadrons, as one expects to see their signature in all three final two-body channels.
%
In the $J/\psi \, \phi$ invariant mass spectrum we study the interplay between the $X(4140)$ and the $X(4160)$ resonances. The $J/\psi \, \Lambda$ mass spectrum may help to identify the strange 
partner of the hidden-charm pentaquark recently observed by the LHCb collaboration, the existence of which has been predicted by a chiral unitary approach. We conclude that this strange pentaquark has a good chance of experimental detection if it is present in the range between $4450-4500$ MeV. Finally, in the $\phi \, \Lambda$ spectrum we expect a contribution from a dynamically generated resonance at around $2160$ MeV, but with the present model parameters there is little chance for its experimental detection.       
\end{abstract}


\maketitle

\section{Introduction}

The purpose of this work is to study the $\Lambda_b \rightarrow  J/\psi \ \phi \ \Lambda$ decay and to demonstrate that this reaction is interesting
as it allows for the possibility of detecting exotic hadrons
in the three different two-body invariant mass spectra. The feasibility of measuring this decay has been recently demonstrated by the CMS collaboration \cite{Sirunyan:2019dwp}, where the ratio of the branching fractions ${\cal B}(\Lambda_b \rightarrow J/\psi \ \phi \ \Lambda)/{\cal B}(\Lambda_b \rightarrow \psi(2S)\ \Lambda)$ is measured to be $(8.26 \pm 0.90 ({\rm stat})\pm 0.68 ({\rm syst})\pm 0.11({\cal B}))\times 10^{-2}$, hence opening the door to obtaining spectral information of final-state pairs as soon as sufficient events are observed.
Let us briefly review first the experimental and then the
theoretical works on exotic hadrons that are relevant to the present study.

In 2008, the $X(4160)$ was observed in the $e^+e^- \rightarrow J/\psi \ D^* \ \bar{D}^*$ reaction by the Belle collaboration \cite{Belle1}. Later, in the years from 2009 to 2014, a series of experimental observations of the $X(4140)$ were reported by collaborations such as CDF \cite{CDF1,CDF2}, LHCb \cite{LHCb}, CMS \cite{CMS} and D0 \cite{D01,D02}. All these studies associated a narrow width to this resonance with the average being around $19$ MeV, as reported by the PDG \cite{PDG2}. However, a more recent measurement of the $B^+ \rightarrow J/\psi \ \phi \ K^+$ reaction at LHCb \cite{exp_X,exp_X_another} in 2017 has brought some surprises. The $X(4140)$ deduced from their analysis has a width around $83$ MeV, substantially larger than those claimed in the former experiments. Another noteworthy point is that they have reported several new states that couple to $J/\psi \ \phi$ - $X(4274),X(4550)$ and $X(4700)$. Finally, it is surprising that they do not report the existence of the $X(4160)$, the state seen earlier in 2008, which is presumably related to the fact that they have associated a large width to the $X(4140)$ instead. Therefore it is now debatable as to whether, in the low $J/\psi \ \phi$ spectrum, there is one broad $X(4140)$ or a narrow $X(4140)$ plus another $X(4160)$, see Ref. \cite{wang}, and this is a theme that will play an important role in this work. 

As for the pentaquark baryons, the LHCb collaboration reported the observation of exotic structures in the $J/\psi \, p$ invariant mass spectrum in 2015 \cite{aaij} while observing the $\Lambda_b^0 \rightarrow J/\psi \ p \ K^-$ decay. The data shows a clear spike at $4.5$ GeV which was identified as the state $P_c(4450)$. Although not apparent from the $m_{J/\psi p}$ distribution alone, the amplitude analysis also requires a second broad $J/\psi \, p$ state in order to obtain a good description of the data. This state peaks at $4.38$ GeV and was identified as $P_c(4380)$.
More recently, in 2019, the LHCb collaboration has released a more detailed report \cite{LHCnew} on the analysis of the $m_{J/\psi \,p}$ spectrum. The earlier peak at $4.38$ GeV is now identified as a narrow pentaquark state $P_c(4312)$. Also the previously reported $P_c(4450)$ structure is now seen as two overlapping peaks, $P_c(4440)$ and $P_c(4457)$.  The minimal quark content of these states is $duuc\Bar{c}$. Since all these states are narrow and below the $\Sigma^+_c \Bar{D}^0$ and $\Sigma^+_c \Bar{D}^{*0}$ $([duc][u\Bar{c}])$ thresholds within plausible hadron-hadron binding energies, they provide the strongest experimental suggestion to date for the existence of bound states of a baryon and a meson. In simple tightly bound pentaquark models, the proximity of these states to baryon-meson thresholds would be coincidental. See Ref. \cite{Guo} for a review of hadronic molecules.

We now summarize the theoretical works on exotic hadrons. As we mentioned previously, before the LHCb report in 2017, it was widely believed that there was a narrow $X(4140)$ ($\Gamma \approx 19$ MeV) and a rather broad $X(4160)$ ($\Gamma \approx 132$ MeV) in the low $J/\psi \ \phi$ spectrum. In those days, the $X(4140)$ was the subject of many theoretical studies that tried to identify it as a $D_s^* \, \Bar{D}_s^*$ molecule \cite{41401,41402,41403,41404} with the preferred quantum numbers being $0^{++}$ and $2^{++}$. It is interesting to note in hindsight that these theoretical works could have associated the $D_s^* \, \Bar{D}_s^*$ structure to the $X(4160)$ instead of the $X(4140)$. Presumably, the fact that light meson channels were not considered in their work rendered the width of the state small and therefore the association to the $X(4140)$ was more natural. Indeed, after the observation of the $X(4160)$ in 2008, the authors of Ref. \cite{X(4160)} obtained an $X$ ($2^{++}$) state at $4169$ MeV, coupling mostly to $D_s^* \ \Bar{D}_s^*$ and it was associated to the $X(4160)$ and not the $X(4140)$. This is because, having considered light meson channels as the dominant ones, they obtained a large width of $139$ MeV and therefore the association to $X(4160)$ was more natural. Finally, with the quantum numbers of the $X(4140)$ now established to be $1^{++}$ \cite{PDG2}, the association of $0^{++}$ and $2^{++}$ states to the $X(4140)$ can no longer be supported, but the idea of the $D_s^* \, \Bar{D}_s^*$ molecule associated to the $X(4160)$ becomes more plausible.

Nevertheless, as mentioned earlier, the recent LHCb report in 2017
states the existence of only one broad $X(4140)$ and there is no $X(4160)$ in their analysis. But the story continues, and the possibility that the low energy region of the $J/\psi \, \phi$ spectrum can be better explained by two close resonances, $X(4140)$ and $X(4160)$, has been raised recently \cite{wang}. The reasoning is as follows: a close examination of the low $J/\psi \, \phi$ invariant masses seems to disclose a cusp at around $4224$ MeV, which is the value of the $D_s^*\,\Bar{D}_s^*$ threshold, strongly suggesting the involvement of a $D_s^*\,\Bar{D}_s^*$ molecule in the reaction. By assuming a combination of a narrow $X(4140)$ plus a $X(4160)$ (dynamically generated as a $D_s^*\,\Bar{D}_s^*$ molecule following Ref. \cite{X(4160)}) instead of a single broad $X(4140)$, the authors of Ref. \cite{wang} provide a much better fit to the data reproducing the cusp at the $D_s^*\,\Bar{D}_s^*$ threshold. Our study of the $X(4140)/X(4160)$ interplay will closely follow such an approach.

We now move to theoretical studies of the pentaquark. It seems likely that the hidden-charm non-strange pentaquark states recently discovered in the $\Lambda_b^0 \rightarrow J/\psi \  p \ K^-$ decay could have a partner in the strangeness $S=-1$ sector.  
Indeed, calculations in Refs. \cite{Wu1,Wu2} generated non-strange and strange pentaquark states dynamically even before the LHCb discovery. Two states with strangeness and hidden-charm with $J^P=3/2^-$ and isospin $I=0$ were found as meson-baryon molecules having pole positions $\sqrt{s}=4368-2.8i$ and $\sqrt{s}=4547-6.4i$. Later, the authors of Ref. \cite{feijoo} suggested the experimental study of the $\Lambda_b \rightarrow J/\psi \ \eta \ \Lambda$ decay which would allow one to look for a peak corresponding to the strange pentaquark in the $J/\psi \,  \Lambda$ mass distribution. Among the two states discussed earlier, they had chosen, as a candidate for the strange pentaquark, the state with the higher energy having a nominal mass and width of around $4550$ MeV and $10$ MeV, respectively. Although the results obtained seemed to be encouraging, this $\Lambda_b \rightarrow J/\psi \ \eta \ \Lambda$ decay has not yet been studied experimentally due to the experimental difficulties in reconstructing the $\eta$ particle. We argue that the $\Lambda_b \rightarrow J/\psi \ \phi \ \Lambda$ decay may provide a chance to observe the strange pentaquark.

We will also explore the possibility that a strange resonance, generated from vector meson-baryon dynamics in  Ref. \cite{oset}
and located below the threshold of $K^* \Xi$ states to which it couples strongly, could contribute a significant peak in the corresponding invariant $\phi\,\Lambda$ mass spectrum.

An outline of the following sections is as follows. In Section 2, we present the Dalitz plots, based on which we will justify the general motivation to study the $\Lambda_b \rightarrow  J/\psi \ \phi \ \Lambda$ decay. In Section 3, we construct the invariant amplitude which is used to calculate the differential decay width as a function of the two-particle invariant masses. In Section 4, we present a discussion of our results and end with the conclusion.

\section{General Motivation: Dalitz Plots}

From the kinematics of a 3-body decay, we can construct Dalitz plots which depict the kinematically available phase space in the reaction. Also, from the experimental and theoretical studies discussed in the introduction, we have an idea of the location of various interesting, new and exotic resonances and we will see that many of them lie inside the kinematically allowed region, making a strong case for the study of the $\Lambda_b \rightarrow J/\psi \ \phi \ \Lambda$ decay. In the Dalitz plots, the energy position of the resonances are drawn as lines of different style, while a color area around each line depicts the corresponding width. 

\begin{figure}
\includegraphics[scale=0.33]{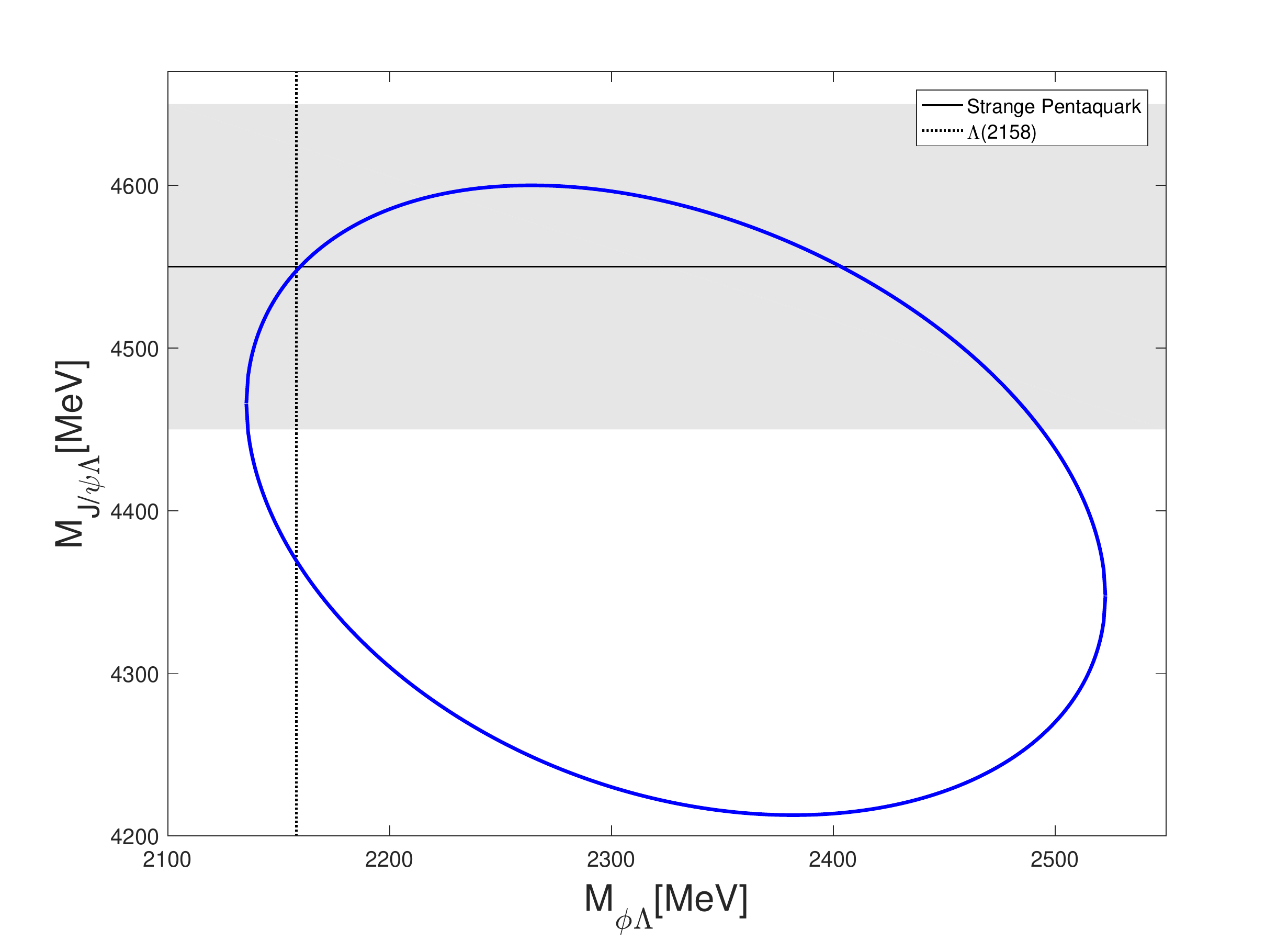}
\caption{(Color online) Dalitz plot for $M_{J\psi\Lambda}$ as a function of $M_{\phi\Lambda}$. The vertical line represents the $\phi\Lambda$ resonance at position $M_{\phi\Lambda}=2158$ MeV with $\Gamma=13$ MeV. The horizontal line is the pentaquark resonance with position $M_{J/\psi\Lambda}=4550$ MeV with $\Gamma=10$ MeV. A possible variation of $100$ MeV has been considered for the position of the pentaquark.}
\label{DP1}
\end{figure}

\begin{figure}
\centering
\includegraphics[scale=0.33]{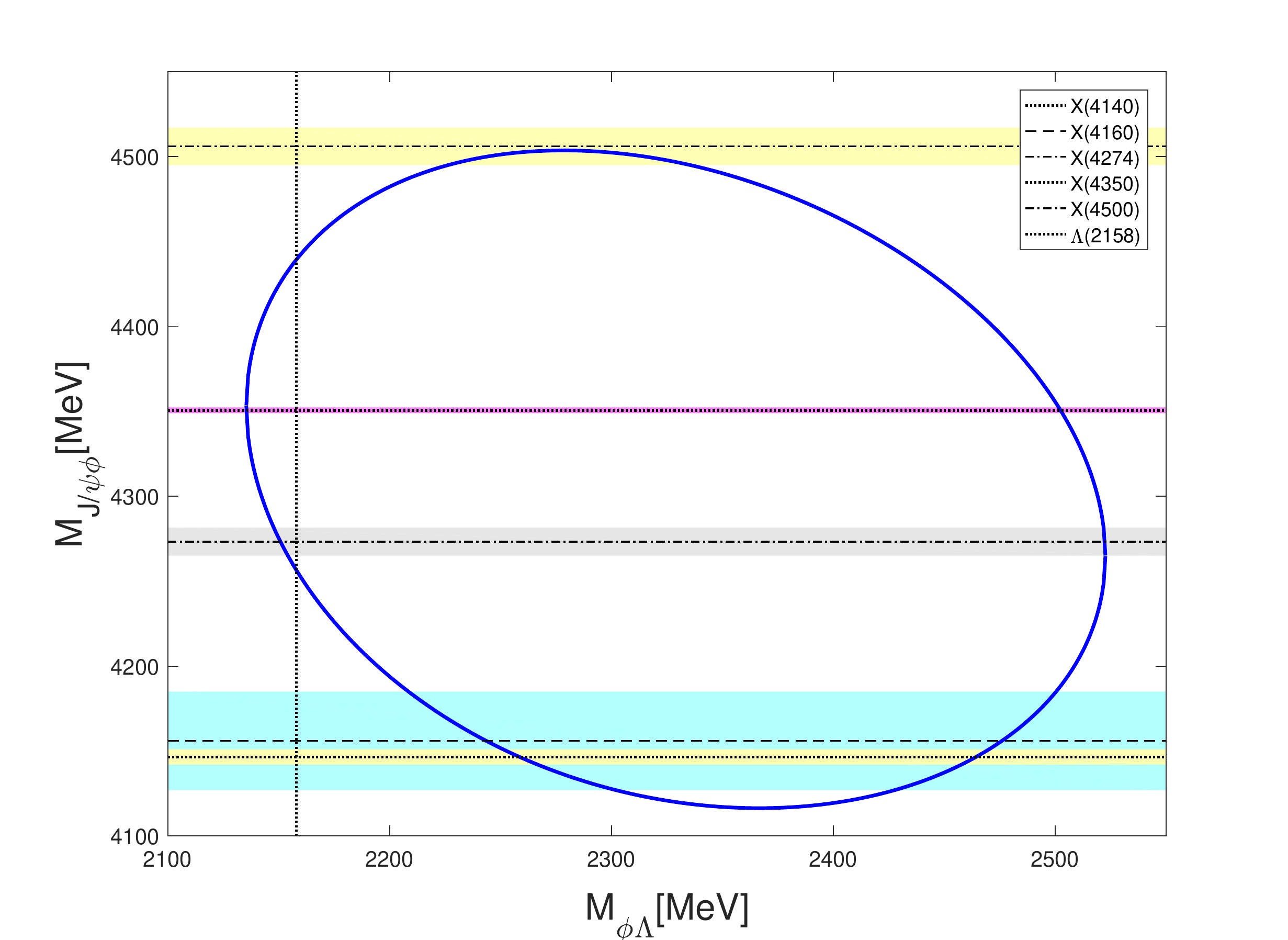}
\caption{(Color online) Dalitz plot for $M_{J/\psi\phi}$ as a function of $M_{\phi \Lambda}$.The resonances corresponding to the $J/\psi \ \phi$ channel are $M_{X(4500)}=4506\pm 11$ MeV with $\Gamma=92$ MeV,  $M_{X(4350)}=4350.6\pm 0.7$ MeV with $\Gamma=13$ MeV,  $M_{X(4274)}=4273.3\pm 8.3$ MeV with $\Gamma=56.2$ MeV,  $M_{X(4160)}=4156\pm 29$ MeV with$\Gamma=139$ MeV,  and $M_{X(4140)}=4146.5\pm 4.5$ MeV with $\Gamma=83$ MeV. The parameters of the $\phi\Lambda$ resonance coincide with the ones used in Fig. \ref{DP1}.}
\label{DP2}
\end{figure}

\begin{figure}
\includegraphics[scale=0.33]{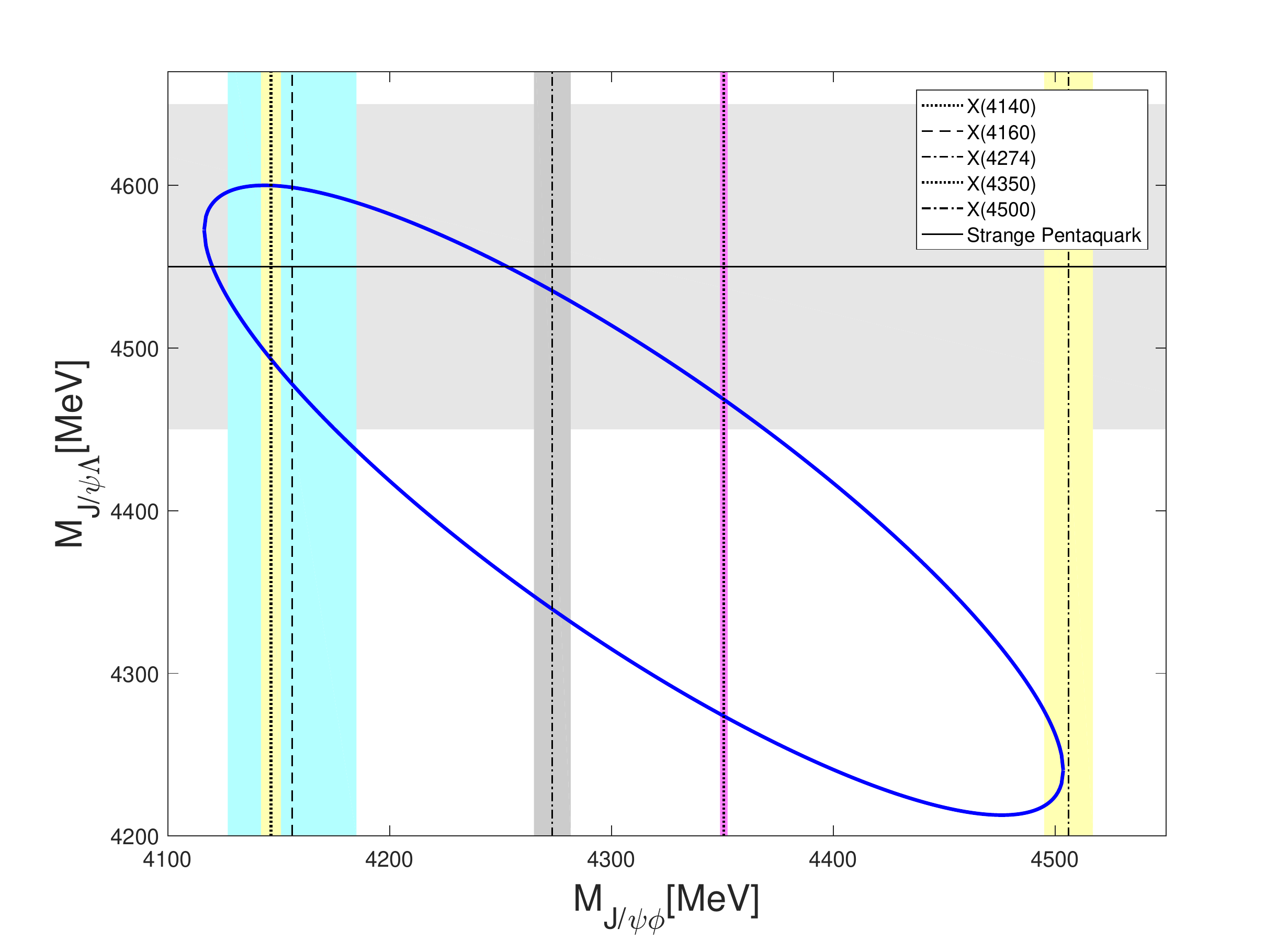}
\caption{(Color online) Dalitz plot for $M_{J/\psi\Lambda}$ as a function of $M_{J/\psi\phi}$. The resonances are the same as in Figs. \ref{DP1} and Fig. \ref{DP2}.}
\label{DP3}
\end{figure}

The $M_{J/ \psi\Lambda}$ vs $M_{\phi\Lambda}$ Dalitz plot is shown in Fig. \ref{DP1}. The resonance theoretically found in Ref. \cite{oset} is inside the $\phi\,\Lambda$ kinematic region of interest and it may contribute a significant peak in the corresponding invariant mass spectrum. The $M_{J/\psi \phi}$ vs $M_{\phi\Lambda}$ Dalitz plot shown in Fig. \ref{DP2} displays several $J/\psi \, \phi$ resonances lying within the kinematically allowed region. These are the $X(4140)$, $X(4160)$, $X(4274)$, $X(4350)$ and $X(4500)$ resonances, although the last one is at the kinematic limit. Among these resonances, there is presently no theoretical model for the last three and therefore we shall not consider them in our study. This doesn't mean that they cannot be observed in the data, only that we at this moment cannot make any credible prediction for their production.  
Finally in Fig. \ref{DP3} we present the Dalitz plot for $M_{J/\psi \Lambda}$ as a function of $M_{J/\psi\phi}$. As in Fig. \ref{DP2}, one can see up-to five $X$ resonances in the $J/\psi \, \phi$ channel. In the $J/\psi \, \Lambda$ channel, we should see the presence of the predictions of Refs. \cite{Wu1,Wu2} where they introduce the idea of a possible strange partner to the hidden-charm pentaquark observed by LHCb \cite{aaij,LHCnew}. Among the two states that have been predicted, we have picked, following Ref. \cite{feijoo}, the one with the larger energy of $4550$ MeV and a width of $10$ MeV.

In summary, in this section we have shown that in every two-particle invariant mass spectra we have the opportunity to observe new and exotic resonances that have been predicted by previous theoretical works or seen in experiments,
thus serving as a general motivation for studying the $\Lambda_b \rightarrow  J/\psi \ \phi \ \Lambda$ decay.

\section{Formalism}
The formalism we have adopted to describe the decay of the $\Lambda_b$ into $J/\psi \ \phi \ \Lambda$ is now discussed. The construction of the amplitude for this decay will reflect some of the various interesting features that can be studied with this reaction. The interactions appearing in the decay model are calculated with chiral Lagrangians \cite{scherer,pich,kaplan,Weinberg} given by the hidden gauge approach \cite{HG1,HG2,HG3} complemented with non-perturbative techniques \cite{Oller,oset_chiral2,oset_chiral1,hyodo,ramos_chiral}.

\subsection{Primary decay mechanism}

\begin{figure}
    \centering
    \includegraphics[scale=0.74]{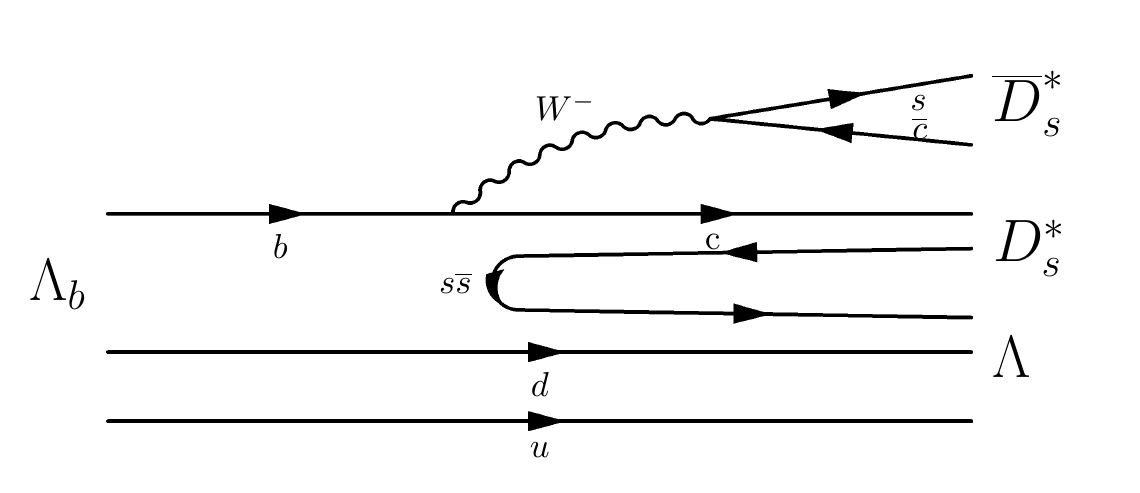}
    \caption{Microscopic quark level production of $\Lambda \ D_s^* \ \Bar{D}_s^*$ in the $\Lambda_b$ decay through external emission.}
    \label{ext_emi}
\end{figure}

One of the motivations for studying the $\Lambda_b \rightarrow J/\psi \ \phi \ \Lambda$ decay is that we can study the interplay between the $X(4140)$ and $X(4160)$ resonances, similar to how it was done in Ref. \cite{wang}. Therefore we consider two separate primary decay mechanisms, one involving the $X(4160)$ and the other involving the $X(4140)$, but these resonances will not be treated on the same footing in this work. The $X(4160)$ is obtained as the result of a chiral unitarized coupled-channels calculation first performed in Ref. \cite{X(4160)} and reproduced here again in this work. On the other hand, the $X(4140)$ is introduced as a Breit-Wigner whose parameters are fit to experimental data. Thus, while the $X(4160)$ has a clear interpretation as a meson-meson molecule, the $X(4140)$ has no such simple physical interpretation and is introduced purely on a phenomenological basis. Taking this into account, we first discuss the production mechanism involving the $X(4160)$.

\begin{figure}
    \centering
\includegraphics[scale=0.96]{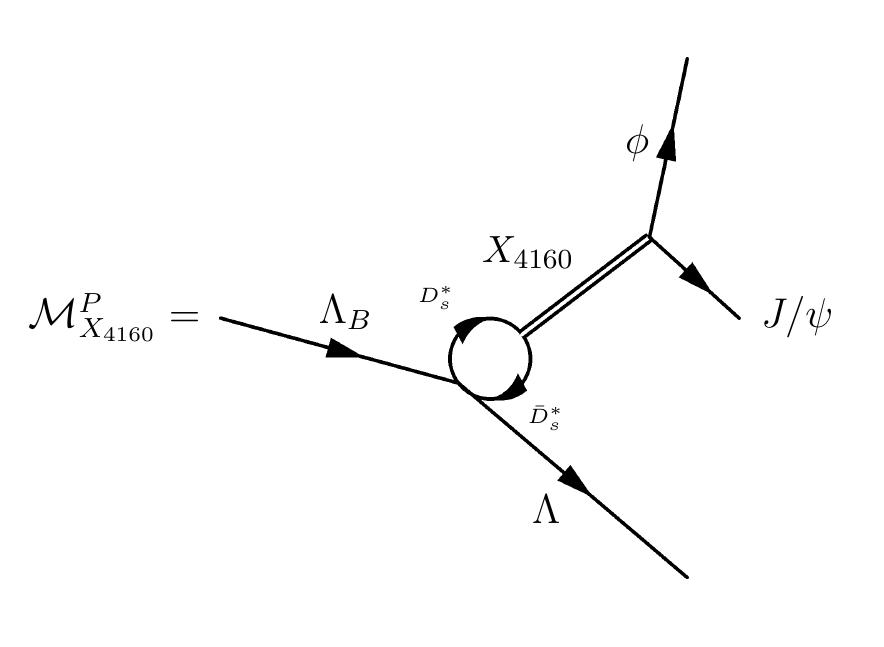}
    \caption{Mechanism for the $\Lambda_b$ decay into $J/\psi \ \phi \ \Lambda$ in the presence of the $X(4160)$ resonance.}
    \label{pr_4160}
\end{figure}

In the coupled-channels approach, the $X(4160)$ has its strongest coupling to the $D_s^* \, \Bar{D}_s^*$ channel. This requires us to discuss how a $D_s^*\, \Bar{D}_s^*$ pair is produced in the weak decay $\Lambda_b \rightarrow J/\psi \ \phi \ \Lambda$. The dominant process at the quark level proceeds via the external emission mechanism depicted in Fig. \ref{ext_emi}. Since the final state involves $J/\psi \,\phi$ pairs,  the $D_s^* \, \Bar{D}_s^*$ lines are closed to form a loop, allowing for multiple interactions in coupled channels that lead to the dynamical generation of the $X(4160)$ which then decays into $J/\psi$ and $\phi$, as shown in Fig. \ref{pr_4160}.
The corresponding amplitude is written as
\begin{equation}
    \mathcal{M}^{P}_{X_{4160}}= A (\Vec{\epsilon}_{J/\psi}\times \Vec{\epsilon}_{\phi}) \cdot \Vec{P}_{\Lambda} \ G_{D_s^* \, \Bar{D}_s^*} \ \frac{T_{D_s^*  \Bar{D}_s^*, J/\psi  \phi}}{g_{D_s^*  \Bar{D}_s^*} g_{J/\psi  \phi}} \ ,
    \label{5.1}
\end{equation}
and stands for the primary production of the $J/\psi \ \phi \ \Lambda$ final state involving the $X(4160)$.

Several comments are in order regarding Eq. (\ref{5.1}). The amplitude is denoted as $ \mathcal{M}^{P}_{X_{4160}}$, where the superscript $P$ refers to ``Primary decay'' and the subscript indicates that the reaction proceeds via the $X(4160)$. On the right-hand side, we first have the constant $A$ which represents the amplitude for the microscopic decay transition at quark level shown in Fig. \ref{ext_emi}. The calculation of $A$ is quite difficult and is well beyond the scope of this work. 
However, for our purposes and similarly as it was argued in Ref. \cite{oset_prc92}, this amplitude can be taken as a constant in the limited range of energies involved in the $\Lambda_b \rightarrow J/\psi \ \phi \ \Lambda$ decay. The factor $G_{D_s^* \, \Bar{D}_s^*}$ stands for the $\ D_s^* \, \Bar{D}_s^*$ loop that appears in Fig.~\ref{pr_4160}. The symbol $T_{D_s^*  \Bar{D}_s^*, J/\psi  \phi}$ denotes the coupled-channel unitarized amplitude for the $D_s^* \, \Bar{D}_s^* \rightarrow J/\psi \, \phi$ process and it is the piece that contains the dynamically generated  $X(4160)$ \footnote{Note that we have divided the amplitude by the coupling constants of the resonance to the meson-baryon states at the production and decay ends. This is done to facilitate the combination of this contribution with that of the $X(4140)$ using a coupling strength $B$ that will have the same units as $A$.}. 
The factor $(\Vec{\epsilon}_{J/\psi}\times \Vec{\epsilon}_{\phi}) \cdot \Vec{P}_{\Lambda}$ is indicative of a P-wave operator which is the minimum partial wave needed to conserve total angular momentum at the weak vertex, noting that the spin quantum number of the $X(4160)$ is $J=2$ while those of the $\Lambda_b$ and $\Lambda$ are $J=1/2$.
Here, $\Vec{\epsilon}_i$ denotes the polarization vector of the corresponding vector meson in the $J/\psi  \, \phi $ rest frame and $\Vec{P}_{\Lambda}$ is the three-momentum of the $\Lambda$ in the same frame. As we will see, working in the $J/\psi \, \phi$ rest frame will allow us to evaluate the required spin sums in an easy manner. 

\begin{figure}
    \centering
    \includegraphics[scale=0.24]{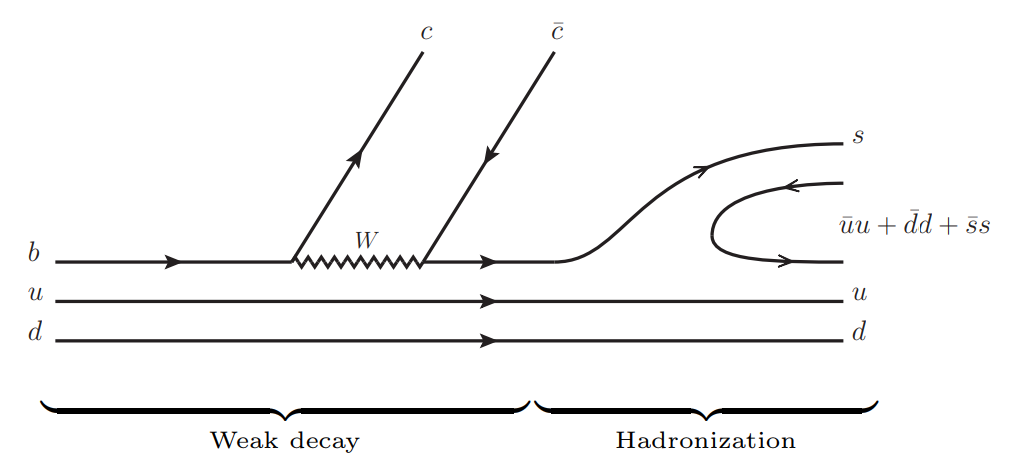}
    \caption{Microscopic quark level production of $J/\psi \ \phi \ \Lambda$ in the $\Lambda_b$ decay through internal conversion.}
    \label{int_con}
\end{figure}

We note that, although the $X(4160)$ couples with maximum strength to the $D_s^* \, \Bar{D}_s^*$ channel, it also evidently couples with significant strength to the $J/\psi \, \phi$ channel. Therefore one could, in principle, have diagrams similar to Fig. \ref{pr_4160} but with a $J/\psi \, \phi$ virtual state instead of $D_s^* \, \Bar{D}_s^*$. But, at the microscopic quark level, this reaction  proceeds via the internal conversion process shown in Fig. \ref{int_con}, which is strongly penalized by color factors \cite{color_fac} and therefore it will be neglected in this work, similar to what was done in Ref. \cite{wang}.

\begin{figure}
    \centering
\includegraphics[scale=0.96]{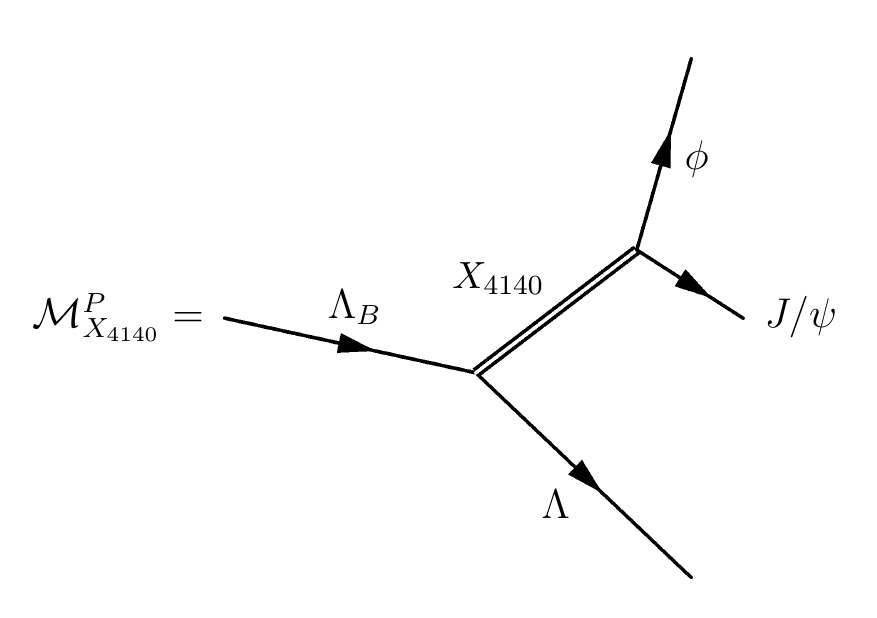}
    \caption{Mechanism for the $\Lambda_b$ decay into $J/\psi \ \phi \ \Lambda$ in the presence of the $X(4140)$ resonance.}
    \label{pr_4140}
\end{figure}

We now turn to the primary production mechanism via the $X(4140)$. The process is shown in Fig. \ref{pr_4140} and the associated amplitude takes the form
\begin{equation}
    \mathcal{M}^{P}_{X_{4140}}= \frac{\Tilde{B} }{2M_{X(4140)}\big[M_{J/\psi \phi}-M_{X(4140)}+i\frac{\displaystyle \Gamma_{X(4140)}}{\displaystyle 2} \big]},
\end{equation}
where $\Tilde{B}$ is a constant connected to the combined strength of the $\Lambda_b \rightarrow \Lambda \ X(4140)$ vertex and that of its subsequent decay into $J/\psi\,\phi$ states. We will write it as $\Tilde{B}=B M_{X(4140)}$ so that $A$ and $B$ have the same units. As will be discussed later, only the ratio between $B$ and $A$ is important and not their absolute values. The $X(4140)$ is parameterised with a Breit-Wigner, where $M_{J/\psi \phi}$ is the invariant mass of the $J/\psi \, \phi$ system, and $M_{X(4140)}$ and $\Gamma_{X(4140)}$ stand for the mass and width of the $X(4140)$, respectively. Their explicit values will be given later when discussing the results. Note that, unlike the previous case of the $X(4160)$, it is not necessary to have a $P-$wave vertex here because the spin quantum number of the $X(4140)$ is $J=1$. Therefore we take it to be $S-$wave.

\begin{figure*}[t]
\centering
\includegraphics[width=0.4\textwidth]{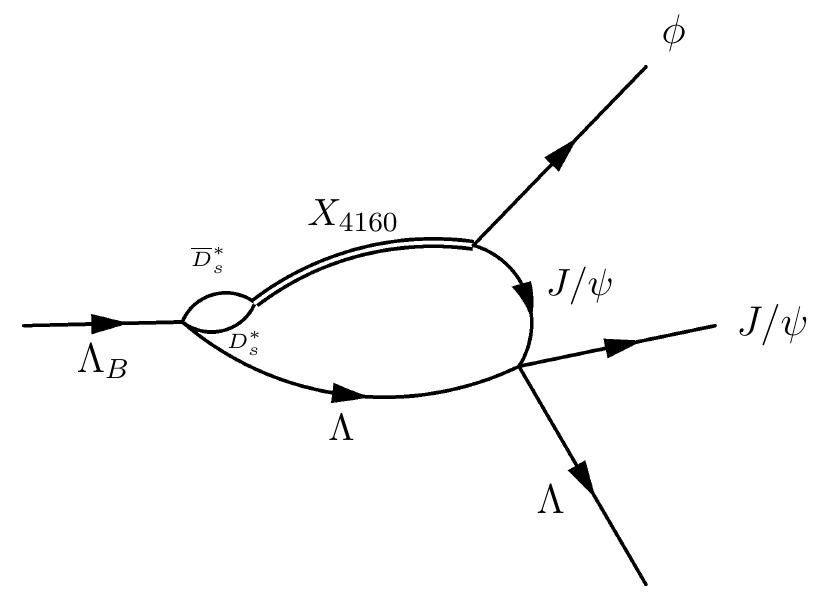}
     \hspace{25pt}
\includegraphics[width=0.4\textwidth]{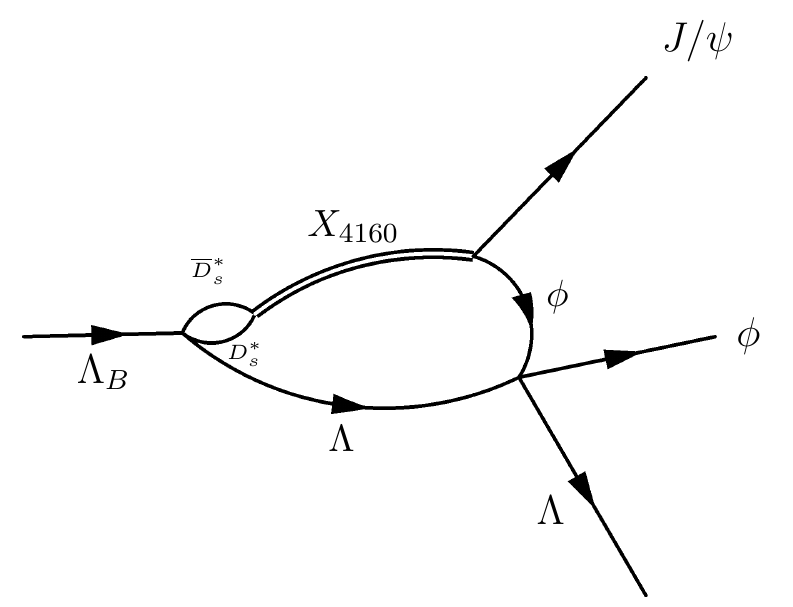}

\hspace{0.1\textwidth} (a)\hspace{0.45\textwidth} (b) \hspace{0.1\textwidth}

\caption{Final state interaction between $J/\psi \ \Lambda$ (left) and $\phi \ \Lambda$ (right) in the presence of the $X(4160)$ resonance.}
\label{fin_state_4160}
\end{figure*}

As commented above, the philosophy of our model is to consider the lowest partial wave that preserves conservation of angular momentum in each contribution. In the case of the X(4160), the weak vertex describing the decay of the $\Lambda_b$ ($J^P=1/2^+$) to the $X(4160)$ ($J^P=2^+$) and the $\Lambda$  ($J^P=1/2^+$)  requires a minimum angular momentum value of $L=1$, therefore leading to a parity violating amplitude. In the case of the $X(4140)$ ($J^P=1^+$) mechanism, the minimum angular momentum is $L=0$ leading to a parity conserving contribution. The consideration of higher angular momenta (hence different parity terms) within each mechanism is out of the scope of the present work. It would involve the introduction of new parameters on which one does not have any information. Nevertheless, we believe that the aim of this work, pointing towards the feasibility of using the decay of the $\Lambda_b$ into $ J/\Psi\ \phi\ \Lambda$ to identify possible exotic states from the analyses of two-body invariant masses in the final state, remains valid within this simplified treatment of the weak vertex.

\subsection{Final State Interaction}

The processes discussed in the previous section allow us to obtain interesting features in the $J/\psi \, \phi$ invariant mass spectrum. However, we would also like to monitor the other two-particle invariant mass spectra. In the $J/\psi \, \Lambda$ spectrum we would like to study the possible traces of the strange pentaquark and, in the $\phi \, \Lambda$ spectrum, a resonance dynamically generated in Ref. \cite{oset}. For these purposes we require final state interaction between $J/\psi \, \Lambda$ pairs and $\phi \, \Lambda$ pairs, which can be implemented starting from the primary production processes shown in the previous section, driven by the $X(4160)$ and $X(4140)$ resonances. The external $J/\psi$ and $\Lambda$ legs of Fig.~\ref{pr_4160} are closed to form a loop which, by virtue of the $J/\psi \, \Lambda$ interaction, can dynamically create a resonance, as shown in Fig. \ref{fin_state_4160}(a). The $\phi \, \Lambda$ interaction is constructed in the same manner [see Fig. \ref{fin_state_4160}(b)].

The analytical expression associated with the diagram for $J/\psi \ \Lambda$ scattering (Fig. \ref{fin_state_4160}(a)) is 
\begin{equation}
   \mathcal{M}^{J/\psi \Lambda}_{X_{4160}} = A (\Vec{\epsilon}_{J/\psi}\times \Vec{\epsilon}_{\phi}) \cdot \bigg(\frac{\Vec{P}_{\Lambda}-\Vec{P}_{\phi}}{2} \bigg) \ \ \  T_{J/\psi \Lambda,J/\psi \Lambda} \  I^{J/\psi \Lambda}_{X_{4160}},
   \label{5.3}
\end{equation}
where we have denoted the amplitude as $\mathcal{M}^{J/\psi \Lambda}_{X_{4160}}$ with the superscript indicating the $J/\psi \, \Lambda$ final state interaction. This expression involves the evaluation of a loop integral with several propagators. It is explicitly derived in Appendix \ref{DoubleLoopIntegrals} and we only sketch here the outline of the calculation. Note that, as in the previous section, we require the weak decay vertex to be a $P-$wave one. Since the corresponding $P-$wave operator assigned to the $\Lambda_b \rightarrow D_s^* \ \Bar{D}_s^* \ \Lambda$ process is proportional to the momentum of $\Lambda$ [see Eq.~(\ref{5.1})], we have to deal with the fact that the loop integral is a three-vector. In order to take into account this issue, we have evaluated the loop integral in the $J/\psi \, \Lambda$ rest frame, where it can be shown (see Appendix \ref{DoubleLoopIntegrals}) that, in the non-relativistic limit, it is proportional to the vector $(\Vec{P}_{\Lambda}-\Vec{P}_{\phi})$ times a scalar loop integral. This explains the momentum and polarization factors appearing in Eq. (\ref{5.3}) with $I^{J/\psi \Lambda}_{X_{4160}}$ being the scalar loop integral.
Finally, $ T_{J/\psi \Lambda,J/\psi \Lambda}$ captures the $J/\psi \  \Lambda$ final state interaction and is given as 
\begin{equation}
    T_{J/\psi \Lambda,J/\psi \Lambda} = \frac{g_{J/\psi \Lambda}^2}{M_{J/\psi \Lambda}-M + i\frac{\Gamma}{2}},
\end{equation}
where $M_{J/\psi \Lambda}$ is the invariant mass of the $J/\psi \, \Lambda$ system. The complex number $g_{J/\psi \Lambda}$ is the coupling of the pentaquark to the $J/\psi \,  \Lambda$ channel. 
Note that, although we have used a Breit-Wigner representation for the pentaquark, the values of its  parameters ($M$, $\Gamma$ and $g_{J/\psi \Lambda}$) are extracted from the behaviour of the amplitude calculated using a chiral, unitary coupled-channels approach \cite{Wu1,Wu2}.

The calculation of the $\phi \, \Lambda$ final state interaction proceeds in exactly the same manner. We have 
\begin{equation}
   \mathcal{M}^{\phi \Lambda}_{X_{4160}} = A (\Vec{\epsilon}_{J/\psi}\times \Vec{\epsilon}_{\phi}) \cdot \bigg(\frac{\Vec{P}_{\Lambda}+\Vec{P}_{\phi}}{2} \bigg) \ \ \  T_{\phi \Lambda,\phi \Lambda} \  I^{\phi \Lambda}_{X_{4160}},
   \label{5.5}
\end{equation}
where $T_{\phi \Lambda,\phi \Lambda}$ is the amplitude for the $\phi \Lambda$ interaction that is calculated in a chiral, unitary coupled-channels approach,  following Ref.~\cite{oset}. The remaining terms are analogous to the ones in Eq. (\ref{5.3}).

\begin{figure*}
\centering
\includegraphics[width=0.4\textwidth]{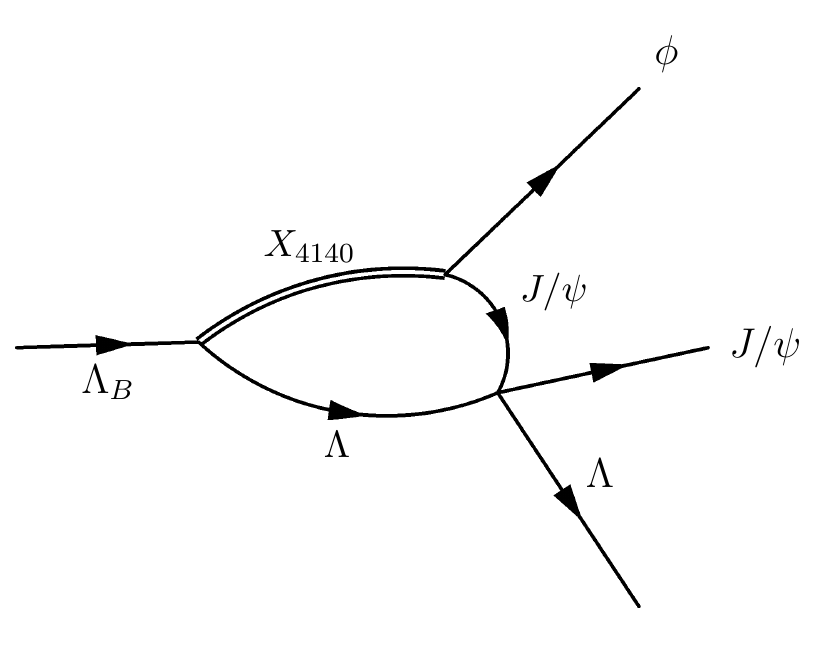}
     \hspace{25pt}
\includegraphics[width=0.4\textwidth]{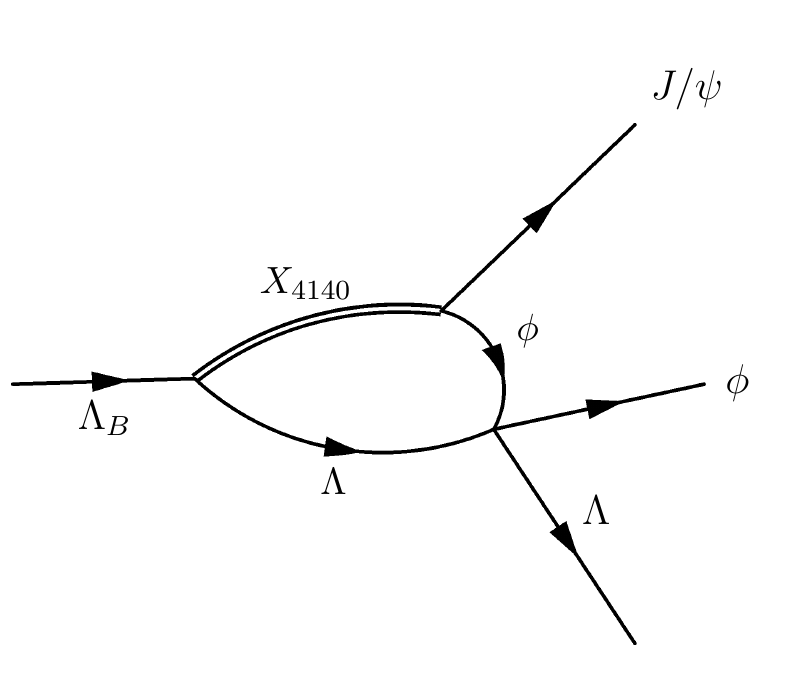}

\hspace{0.1\textwidth} (a)\hspace{0.45\textwidth} (b) \hspace{0.1\textwidth}

     \caption{Final state interaction between $J/\psi \Lambda$ (left) and $\phi \Lambda$ (right) in the presence of the $X(4140)$ resonance.}
 \label{fin_state_4140}
\end{figure*}


So far, we have concentrated on final state interactions in the presence of the $X(4160)$. We can also obtain final state interaction diagrams similar to Figs. \ref{fin_state_4160}(a) and \ref{fin_state_4160}(b) but now starting from the primary process involving the $X(4140)$ and these are shown in Figs. \ref{fin_state_4140}(a) and \ref{fin_state_4140}(b). The associated analytic expressions are
\begin{equation}
    \mathcal{M}^{J/\psi \Lambda}_{X_{4140}}= \Tilde{B} \ 
    T_{J/\psi \Lambda,J/\psi \Lambda} I^{J/\psi \Lambda}_{X_{4140}}, 
\end{equation}
\begin{equation}
   \mathcal{M}^{\phi \Lambda}_{X_{4140}}= \Tilde{B} \  
   T_{\phi \Lambda,\phi \Lambda} I^{\phi \Lambda}_{X_{4140}}.  
\end{equation}
The terms $I^{J/\psi \Lambda}_{X_{4140}}$ and $I^{\phi \Lambda}_{X_{4140}}$ are scalar loop integrals analogous to the ones in Eqs. (\ref{5.3}) and (\ref{5.5}). Their expressions are given in Appendix \ref{DoubleLoopIntegrals}. 

\subsection{The Full Amplitude}

In order to proceed, we now have to combine the various terms to create the full, invariant amplitude. It will turn out to be useful to first collect the contributions
corresponding only to $X(4160)$ as $\mathcal{M}_{X_{4160}}$ and similarly for $X(4140)$,
\begin{equation}
    \mathcal{M}_{X_{4160}} = \mathcal{M}^{P}_{X_{4160}} + \mathcal{M}^{J/\psi \Lambda}_{X_{4160}} + \mathcal{M}^{\phi \Lambda}_{X_{4160}},
\end{equation}
\begin{equation}
    \mathcal{M}_{X_{4140}} = \mathcal{M}^{P}_{X_{4140}} + \mathcal{M}^{J/\psi \Lambda}_{X_{4140}} + \mathcal{M}^{\phi \Lambda}_{X_{4140}}.
    \label{5.9}
\end{equation}
Denoting the full amplitude as $\mathcal{M}$, we have
\begin{equation}
    \overline{|\mathcal{M}|^2} = \overline{|\mathcal{M}_{X_{4160}}|^2} +  \overline{|\mathcal{M}_{X_{4140}}|^2},
    \label{5.10}
\end{equation}
where the bar represents a sum over polarizations. The reason that we have an incoherent sum, $\overline{|\mathcal{M}_{X_{4160}}|^2} +  \overline{|\mathcal{M}_{X_{4140}}|^2}$, and not $\overline{|\mathcal{M}_{X_{4160}} + \mathcal{M}_{X_{4140}}|^2}$ is because the weak decay vertex is $P-$wave in the $X(4160)$ contribution while it is $S-$wave in the case of the $X(4140)$ and these two orthogonal partial waves do not interfere. 

Before proceeding further, we implement the useful redefinitions
\begin{equation}
    \mathcal{M}_{X_{4160}} \rightarrow A\, \mathcal{M}_{X_{4160}} \ ,
\end{equation}
\begin{equation}
    \mathcal{M}_{X_{4140}} \rightarrow B\, \mathcal{M}_{X_{4140}} \ ,
\end{equation}
such that
Eq. (\ref{5.10}) now becomes
\begin{equation}
\begin{split}
      \overline{|\mathcal{M}|^2} &= |A|^2 \ \overline{|\mathcal{M}_{X_{4160}}|^2} + |B|^2 \ \overline{|\mathcal{M}_{X_{4140}}|^2} \\ 
      &\quad = |A|^2 \bigg(\overline{|\mathcal{M}_{X_{4160}}|^2} + \beta \ \overline{|\mathcal{M}_{X_{4140}}|^2} \bigg),
\end{split}
\end{equation}
where we have defined $\beta \equiv |B|^2/|A|^2$. The overall factor of $|A|^2$ is not important since all our final plots will be in arbitrary units. Therefore it will be set to $1$ from now on, but $\beta$ is an important parameter that acts as the relative weight between the $X(4140)$ and the $X(4160)$ contributions. We do not have any means of calculating it directly and also we do not have data to which this parameter can be fit. To solve this issue, we take the value of $\beta$ from Ref. \cite{wang} where the authors have studied the $X(4140)/X(4160)$ interplay in the context of the decay $B^+ \rightarrow J/\psi \ \phi \ K^+$ and we justify this as follows. For the reaction mechanism involving the $X(4160)$ the $B^+ \rightarrow J/\psi \ \phi \ K^+$ decay at the microscopic quark level proceeds in the same manner as shown in Fig. \ref{ext_emi}, but without the spectator $d$ quark in the initial and final states. Therefore, the topology of the diagrams at the hadronic level is similar to the ones we use in this work. The latter statement is also true for the case of the $X(4140)$, although there is no underlying microscopic physics for this case. Ultimately, our 
expressions for $\mathcal{M}_{X_{4160}}$ and $\mathcal{M}_{X_{4140}}$ amplitudes are very similar to those of Ref.~\cite{wang}. However, there are a few differences like, for instance, the fact that the weak decay vertex of the process $B^+ \rightarrow J/\psi \ \phi \ K^+$ considered in Ref. \cite{wang} involves partial waves one unity higher than those considered in our approach. Hence, one cannot expect the values of $\beta$ to match exactly and so we should allow for reasonable variations.

We end this section by indicating how to perform the sum over polarization appearing in Eq. (\ref{5.10}). For the S-wave amplitude $\mathcal{M}_{X_{4140}}$, the sum is trivial and it only produces a constant factor that can be absorbed into $\beta$. But, for the P-wave amplitude $\mathcal{M}_{X_{4160}}$ the process is more complicated and we give here only the results, relegating the full calculation to Appendix \ref{SpinSums}. First, upon comparing with Eqs.~(\ref{5.1}), (\ref{5.3}) and (\ref{5.5}), we make the useful definitions:
\begin{equation}
      \mathcal{M}^{P}_{X_{4160}} \equiv (\Vec{\epsilon}_{J/\psi}\times \Vec{\epsilon}_{\phi}) \cdot \Vec{P}_{\Lambda} \ \  \mathcal{\Tilde{M}}^{P}_{X_{4160}},
\end{equation}
\begin{equation}
     \mathcal{M}^{J/\psi \Lambda}_{X_{4160}} \equiv  (\Vec{\epsilon}_{J/\psi}\times \Vec{\epsilon}_{\phi}) \cdot \Vec{K}_2  \ \  \mathcal{\Tilde{M}}^{J/\psi \Lambda}_{X_{4160}}, 
\end{equation}
\begin{equation}
      \mathcal{M}^{\phi \Lambda}_{X_{4160}} \equiv  (\Vec{\epsilon}_{J/\psi}\times \Vec{\epsilon}_{\phi}) \cdot \Vec{K}_1 \ \  \mathcal{\Tilde{M}}^{\phi \Lambda}_{X_{4160}} \ , 
\end{equation}
in terms of the modified amplitudes (denoted with a tilde), where we have defined $\Vec{K}_2 \equiv (\Vec{P}_{\Lambda}-\Vec{P}_{\phi})/2$ and $\Vec{K}_1 \equiv (\Vec{P}_{\Lambda}+\Vec{P}_{\phi})/2$.
Then, it can be shown (see Appendix \ref{SpinSums}) that summing over the $J/\psi$ and the $\phi$ polarizations leads to
\begin{widetext}
\begin{equation}
\begin{split}
    \overline{|\mathcal{M}_{4160}|^2} &= |\Vec{P}_{\Lambda}|^2 |  \mathcal{\Tilde{M}}^{P}_{X_{4160}}|^2 + |\Vec{K}_2|^2 | \mathcal{\Tilde{M}}^{J/\psi \Lambda}_{X_{4160}}|^2  + |\Vec{K}_1|^2 |\mathcal{\Tilde{M}}^{\phi \Lambda}_{X_{4160}}|^2  + 2 \Vec{P}_{\Lambda} \cdot \Vec{K}_2 \  \Re(  \mathcal{\Tilde{M}}^{P}_{X_{4160}}  \mathcal{\Tilde{M}}^{* J/\psi \Lambda}_{X_{4160}})
    \\ 
    &\quad + 2 \Vec{P}_{\Lambda} \cdot \Vec{K}_1 \  \Re(  \mathcal{\Tilde{M}}^{P}_{X_{4160}} \mathcal{\Tilde{M}}^{* \phi \Lambda}_{X_{4160}})  + 2 \Vec{K}_1 \cdot \Vec{K}_2 \  \Re( \mathcal{\Tilde{M}}^{J/\psi \Lambda}_{X_{4160}} \mathcal{\Tilde{M}}^{* \phi \Lambda}_{X_{4160}}),
\end{split}
\label{5.17}
\end{equation}
\end{widetext}
where $\Re$ denotes the real part of the complex argument.

\section{Results}

\subsection{The $J/\psi \ \phi$ Mass Distribution}
We start showing the results for the $J/\psi \ \phi$ mass distribution, which we employ to study the $X(4140)/X(4160)$ interplay. The latest experimental result suggests the presence of a single, broad $X(4140)$ with pole position $M_R=4146.5$ MeV and width $\Gamma = 83$ MeV \cite{exp_X,exp_X_another}. However, earlier experiments \cite{CDF1,CMS,D01} and a recent theoretical study \cite{wang} support the idea of a narrow $X(4140)$ \textit{plus} a wide $X(4160)$. We will study this issue step-by-step by first considering a single broad $X(4140)$ and later adding the $X(4160)$ to a narrow $X(4140)$.

\begin{figure}
\centering
\includegraphics[scale=0.3]{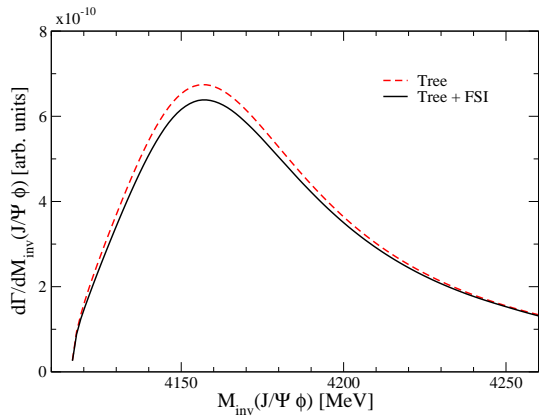}
\caption{$J/\psi \ \phi$ mass spectrum calculated using a broad $X(4140)$ resonance ($M_R=4146.5$ MeV, $\Gamma=83$ MeV). The solid black curve is different from the dashed red curve in that it also includes the effects of the final state interaction.}
\label{J_psi_phi_one}
\end{figure}

Our results for the $J/\psi \, \phi$ spectrum calculated using a single, broad resonance are shown in Fig.~ \ref{J_psi_phi_one}. The red dashed curve corresponds to only the tree level diagram (Fig. \ref{pr_4140}), while the black solid curve takes into account also the final state interaction (Figs. \ref{fin_state_4140}(a) and \ref{fin_state_4140}(b)). The spectrum shows no other noticeable feature than a peak at the pole position of the $X(4140)$'s Breit-Wigner employed in the model. The small difference between the dashed and solid curves indicates the limited effect of final state interactions between the $J\psi \, \Lambda$ and $\phi \, \Lambda$ pairs. We have checked that the same is true for the case of two resonances. Therefore, in the remainder of this section, we shall show only the results obtained including the final state interaction terms in the computation of the $J/\psi \, \phi$ spectrum.  

\begin{figure}
\centering
\includegraphics[scale=0.3]{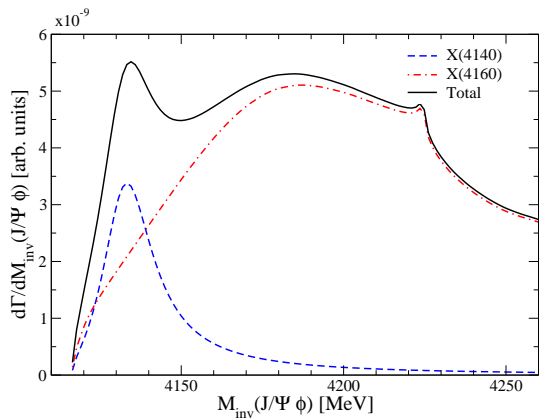}
\caption{The $J/\psi \ \phi$ mass spectrum with a momentum cut-off of $630$ MeV used to regulate $G_{D_s^*\Bar{D}_s^*}$. Individual contributions due to the narrow $X(4140)$ and the $X(4160)$ are shown in  dashed blue and dot-dashed red curves respectively while the full calculation is represented by the solid black line.}
\label{J_psi_phi}
\end{figure}

We now consider calculations done with a narrow $X(4140)$ plus a $X(4160)$. The relative weight between the two resonances $\beta$ is taken from Ref. \cite{wang} but adjusted in order to give roughly equal strengths to the $X(4140)$ and the $X(4160)$ resonances, similar to the results obtained in Ref. \cite{wang}. The role played by $\beta$ will be studied in greater detail later. The pole position and width of the narrow $X(4140)$'s Breit-Wigner are $M_R=4132$ MeV and $\Gamma=19$ MeV. Unlike the $X(4140)$, the $X(4160)$ has the interpretation of a $D_s^*\,\Bar{D}_s^*$ molecule and thus, as discussed, the processes involving the $X(4160)$ contain the loop function $G_{D_s^*\Bar{D}_s^*}$ the divergence of which must be handled by either dimensional regularization (DR) or a momentum cut-off. Even though dimensional regularization was used to compute the unitarized, coupled-channels amplitude that generates the $X(4160)$ \cite{X(4160)}, we do not use it here due to the issues raised in Ref. \cite{loop_prob}, namely the fact that it might lead to  positive values for the real part of the loop that would result in poles in the unitarized amplitude for repulsive potentials, which is unphysical.
We therefore use a cut-off of natural value $630$ MeV which reproduces the loop function calculated with dimensional regularization,   $G_{D_s^*\Bar{D}_s^*}^{DR}$, at the $D_s^*\,\Bar{D}_s^*$ threshold. The results are shown in Fig.~\ref{J_psi_phi} in which the blue dashed and the red dot-dashed curves represent the individual contributions of the $X(4140)$ and the $X(4160)$, respectively, while the black solid curve represents the full calculation which is a weighted sum of the $X(4140)$ and the $X(4160)$.

As we can see, the $X(4140)$ resonance contributes by a peak observed around $4135$ MeV while the $X(4160)$, responsible for most of the remaining strength, produces a broader peak around $4180$ MeV. Finally, we observe a remarkable cusp at the $D_s^*\,\Bar{D}_s^*$ threshold (around $4224$ MeV) which comes from the loop function $G_{D_s^*\,\Bar{D}_s^*}$. We stress that this factor appears as a consequence of analyticity and treating the $X(4160)$ as a $D_s^*\Bar{D}_s^*$ molecule in the coupled-channels approach. Therefore, if such a cusp is observed experimentally, it would not only settle the question as to whether there is one ($X(4140)$) or two resonances ($X(4140)/X(4160)$) but would also strongly suggest a molecular interpretation for the $X(4160)$. A similar result and conclusion was reported in Ref. \cite{wang} where the authors were also able to provide a comparison of their results with experimental data of the $B^+ \rightarrow J/\psi \ \phi \ K^+$ decay.

\begin{figure}
    \centering
        \includegraphics[scale=0.3]{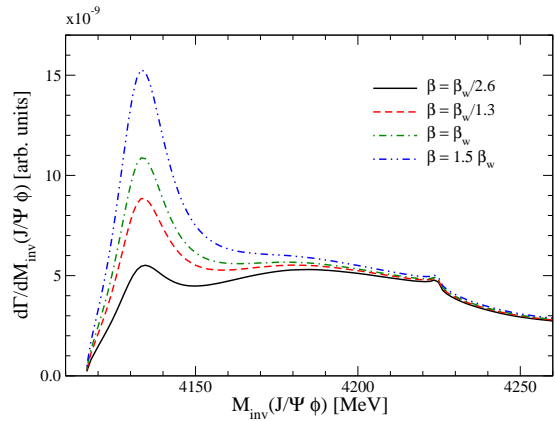}
    \caption{Same as Fig. \ref{J_psi_phi} but now considering different values for the weight factor $\beta$. Here, $\beta_w$ is the value used in Ref. \cite{wang}.}
    \label{diff_beta}
\end{figure}

We end this section with a final analysis regarding the role of the weight factor $\beta$ between the $X(4140)$ and $X(4160)$ contributions. As argued before, we do not expect to obtain a similar invariant mass distribution as that of the $B^- \rightarrow J/\psi \ \phi \ K^-$ decay shown in Fig. 5 of Ref. \cite{wang} for the value of $\beta$ employed there, which we call $\beta_w$. Indeed, we obtain the similarity 
with $\beta=\beta_w/2.6$, which is the value employed in our results of Fig.~\ref{J_psi_phi}. In Fig.~\ref{diff_beta}, we display the spectra obtained for different values for $\beta$, namely $\beta= \beta_w/2.6, \beta_w/1.3, \beta_w$ and $1.5\beta_w$. The most evident difference between the various curves is the height of the $X(4140)$ peak. We also see that the bump corresponding to the $X(4160)$ looses its discernibility as $\beta$ increases but, more importantly, we have the presence of the cusp at the $D_s^*\Bar{D}_s^*$ threshold for all considered values of $\beta$. Although, in an experiment, it could be difficult to notice this cusp if $\beta$ takes values larger than $1.5\beta_w$, this work should serve as a motivation to include this cusp in the experimental amplitude analysis which might lead to a better fit to the data.

\subsection{The $J/\psi \ \Lambda$ Mass Distribution}

In the $J/\psi \ \Lambda$ mass distribution  we expect the signature of the strange pentaquark, whose Breit-Wigner is explicitly put into the decay model, via the final state interaction of $J/\psi \, \Lambda$ pairs. The pole position, width and coupling to the $J/\psi\,\Lambda$ channel are $M_R=4550$ MeV, $\Gamma=10$ MeV and $g_{J/\psi \Lambda}=-0.61-0.06i$ respectively. As mentioned before, these parameters are extracted from the behaviour of the coupled-channels amplitude calculated in Refs. \cite{Wu1,Wu2}. We take these values to be nominal ones and we will explore reasonable variations of them in what follows. 
\begin{figure}
    \centering
    \includegraphics[scale=0.36]{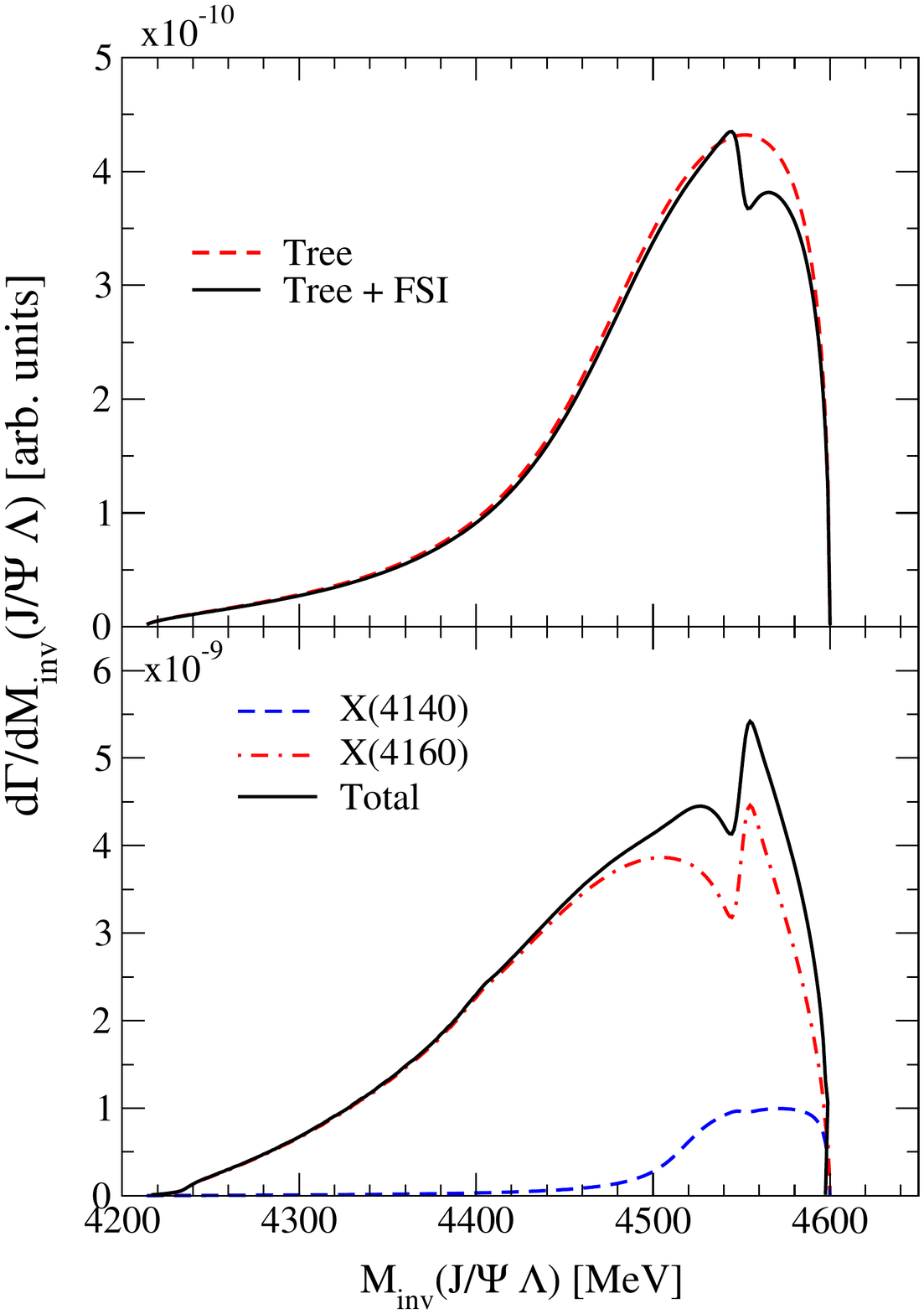}
    \caption{Top panel: The $J/\psi \ \Lambda$ spectrum computed with one broad $X(4140)$. The dashed red curve corresponds to only the tree level diagram while the solid black one takes into account the final state interactions responsible for the pentaquark's generation. Bottom panel: The $J/\psi \ \Lambda$ spectrum computed with a narrow $X(4140)$ plus a $X(4160)$. Individual contributions due to $X(4140)$ and $X(4160)$ are shown in dashed blue and dash-dotted red lines, respectively, while the full calculation is represented by the solid black line. }
    \label{J_psi_lambda_one}
\end{figure}

Similar to our approach in the previous section, we consider two different calculations, one in the presence of a single, broad $X(4140)$ and one in the presence of a narrow $X(4140)$ plus a $X(4160)$. The results for the former case are shown in the upper panel of Fig. \ref{J_psi_lambda_one}, where the red dashed curve represents the background obtained with only the tree-level diagram (Fig. \ref{pr_4140}), while the black solid curve takes into account the final state interaction that proceeds via loop-level processes (Figs. \ref{fin_state_4140}(a) and \ref{fin_state_4140}(b)). Evidently, the solid curve contains the signal of the pentaquark and we must focus on its prominence with respect to the background of the dashed curve. Note that the background shows a broad peak like structure with a maximum located by coincidence roughly at the position of the pentaquark. This can be understood from the $M_{J/\psi \Lambda}$ versus $M_{J/\psi \phi}$ Dalitz plot (Fig. \ref{DP3}) where we see that, around the position of the pentaquark, the range of integration over the $M_{J/\psi \phi}$ invariant masses covers mostly the dominant strength of the $X(4140)$ resonance.  The black solid curve shows a dip at $4550$ MeV. This is the signal corresponding to the pentaquark and it interferes negatively with the background phase space.

Let us next consider the other calculation containing both the $X(4140)$ and the $X(4160)$ resonances, the results of which are shown in the bottom panel of Fig. \ref{J_psi_lambda_one}. The red dot-dashed and the blue dashed curves represent the individual contributions (including the final state interactions) from the $X(4160)$ and the (narrow) $X(4140)$ resonances, respectively, while the black solid curve is the sum of the two with the appropriate weight factor. We notice that the $X(4160)$ is much more dominant than the $X(4140)$. Similar to the previous case where we had only the $X(4140)$, the pentaquark's signal interferes negatively with the background but, quite interestingly, the signals corresponding to these two cases show qualitative differences meaning that the exact features of the low energy region of the $J/\psi \, \phi$ spectrum (one versus two $X$ resonances) have a visible effect on the $J/\psi \, \Lambda$ distribution. This implies that the study of the latter spectrum could help further in learning about the nature of the $X(4140)$ and $X(4160)$. Therefore, in further investigations of the structure of the pentaquark presented in the remainder of this section, we shall show calculations done with one broad $X(4140)$ resonance jointly with those done in the presence of both the $X(4140)$ and $X(4160)$. 

\begin{figure}
     \centering
\includegraphics[scale=0.32]{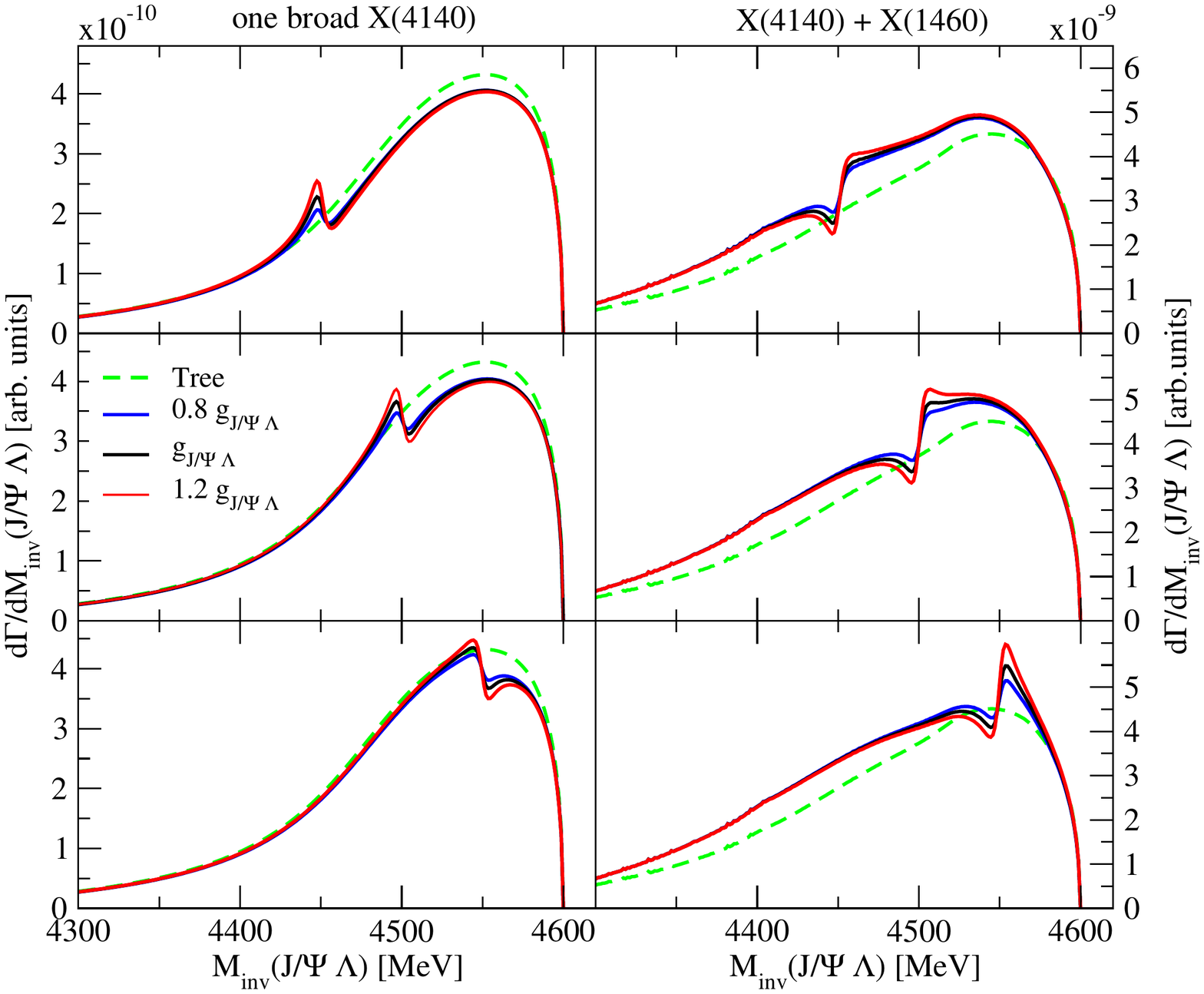}
\caption{The $J/\psi \ \Lambda$ spectrum computed with one broad $X(4140)$ resonance (left panels) or with a narrow $X(4140)$ plus a $X(4160)$ (right panels). Different couplings ($0.8g_{J/\psi \Lambda}$, $g_{J/\psi \Lambda}$, $1.2g_{J/\psi \Lambda}$) and pole positions [$4450$ MeV (top), $4500$ MeV (middle), $4550$ MeV (bottom)] are considered.}
     \label{J_psi_lambda_variations}
\end{figure}

As for the strength of the signal of the pentaquark with respect to the background, we see that it is reasonably sizable to be detected in an experiment. However, before drawing a final conclusion, we explore variations in the parameters of the pentaquark's Breit-Wigner, since one has to accept some uncertainties in the theoretically determined values. In Fig. \ref{J_psi_lambda_variations} we present the calculations done for different values of the pole position of the pentaquark, namely $4450$, $4500$ and $4550$ MeV in the top, middle and bottom panels, respectively. Also, calculations are shown for both cases: one broad resoannce (left panels) versus two $X$ states (right panels). In the former case, if the mass of the pentaquark is $4450$ MeV or lower, it interferes mostly in a positive manner with the background, while for $4550$ MeV or higher masses of the pentaquark the interference is mostly negative. However, in the case where we have two $X$ resonances in the $J/\psi \ \phi$ spectrum, the interference is always negative regardless of the mass of the pentaquark. Furthermore, for all the considered masses of pentaquark, we clearly see the qualitative differences in the $J/\psi \ \Lambda$ spectrum induced by the different cases we are employing to explain the $J/\psi \ \phi$ distribution. Finally, at each pole position we have also considered different couplings of the pentaquark to the $J/\psi \, \Lambda$ channel, namely $0.8g_{J/\psi \Lambda}$, $g_{J/\psi \Lambda}$ and $1.2g_{J/\psi \Lambda}$.  All the considered cases are quite encouraging as the corresponding peak of the pentaquark is discernible over the background, being obviously enhanced for larger values of the coupling strength.

\begin{figure}
    \centering
 \includegraphics[scale=0.32]{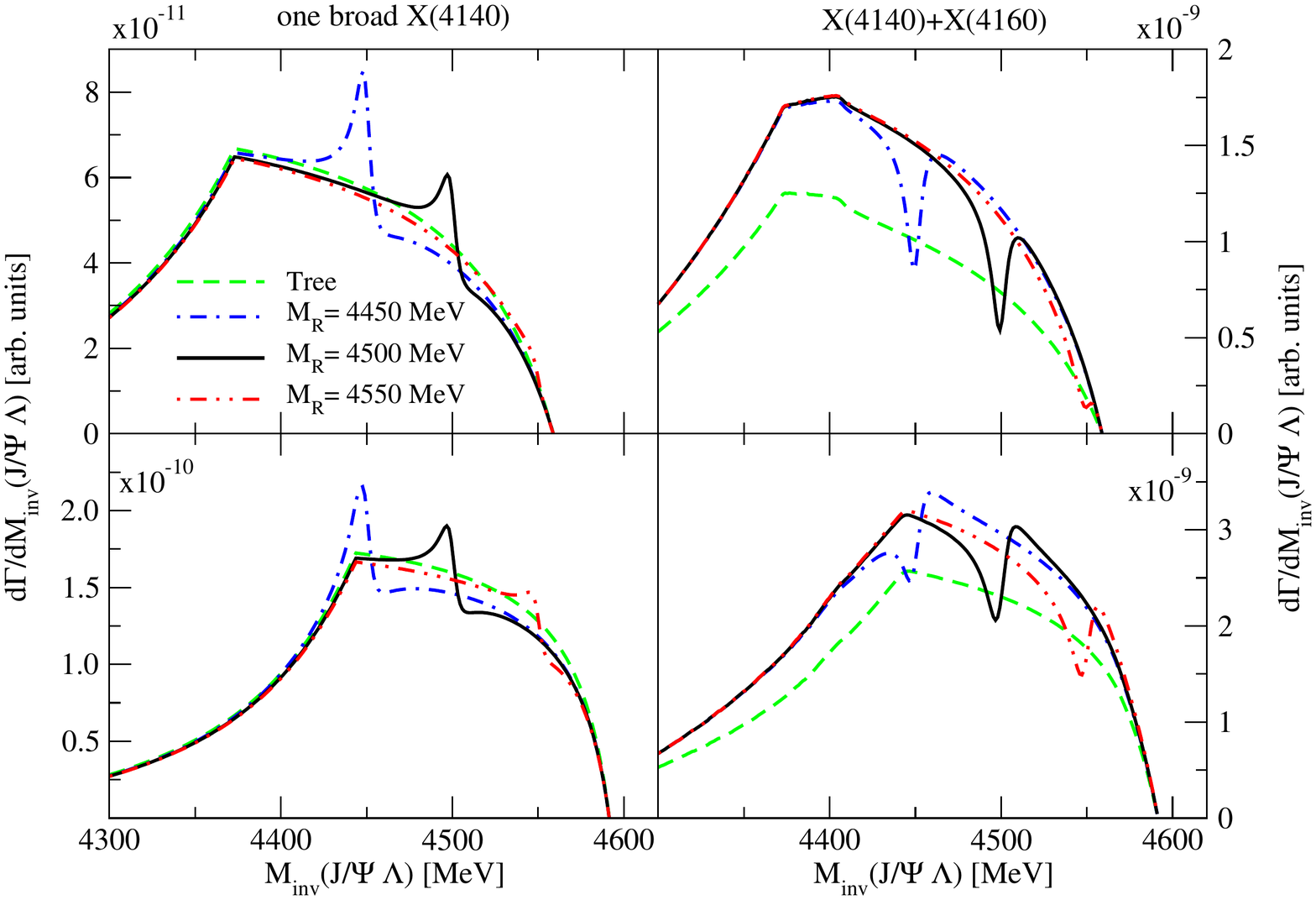}
    \caption{The $J/\psi \ \Lambda$ spectrum computed with one broad $X(4140)$ resonance (left panels) or with a narrow $X(4140)$ plus a $X(4160)$ (right panels), employing the constraints $M_{J/\psi \phi} > 4240$ MeV (top) and $M_{J/\psi \phi}>4180$ MeV (bottom). In each figure, different pole positions for the pentaquark are considered. }
    \label{backcut_penta}
\end{figure}

However the above pentaquark signature can be improved by imposing some cuts in the analysis of the spectra. To illustrate this idea, we present calculations where we impose constraints on the $J/\psi \, \phi$ invariant mass in order to reduce strength coming from the dominant $X$ resonance(s) (either a broad $X(4140)$ or a narrow $X(4140)$ plus a broad $X(4160)$), thereby
enhancing the signal-to-background ratio of the pentaquark. A glance at the $J/\psi \, \Lambda$ vs $J/\psi \, \phi$ Dalitz plot of Fig.~\ref{DP3} shows that cutting the strength of the $X$ resonance(s) while retaining the phase space of the pentaquark is a difficult task and we consider two lower limits on $M_{J/\psi \phi}$. Fig. \ref{backcut_penta} shows the results of these calculations under the conditions $M_{J/\psi \phi}>4240$ MeV (top panels) and $M_{J/\psi \phi} > 4180$ MeV (bottom panels). Different pole positions for the pentaquark are considered, namely $4450,4500$ and $4550$ MeV, which are shown by the dot-dashed blue, solid black and double-dot-dashed red curves, respectively, while the dashed green curve represents the background. The cusp is of no significant interest and arises purely because of the constraints we have imposed on $M_{J/\psi \phi}$. First, let us discuss the top panels where we see that the conditions we have placed on $M_{J/\psi \phi}$ greatly help with the prominence of the pentaquark signal if it has a mass around $4450$ MeV but it has a very destructive effect on the larger masses around $4550$ MeV. This is because if we place a lower bound on $M_{J/\psi \phi}$ that is too large it also kills the phase space of the pentaquark. To correct for this, in the bottom panel we have considered a smaller lower bound on $M_{J/\psi \phi}$, however this does not improve significantly the pentaquark signal at $4550$ MeV. Finally, note that the signal of the pentaquark has a different appearance depending on whether we consider one (left panels) or two $X$ resonances (right panels) in the $J/\psi \ \phi$ spectrum. In the former case we have positive interference leading to peaks while in the latter case we have negative interference leading to dips. 

To summarize all the results of this section, a pentaquark with positions in the range $4450-4500$ MeV (and perhaps with couplings slightly larger than the one predicted in Refs. \cite{Wu1,Wu2}) has a good chance of experimental detection but this deteriorates quickly with increasing values of its mass.

\subsection{The $\phi \ \Lambda$ Mass Distribution}

\begin{figure}[t]
    \centering
    \includegraphics[scale=0.36]{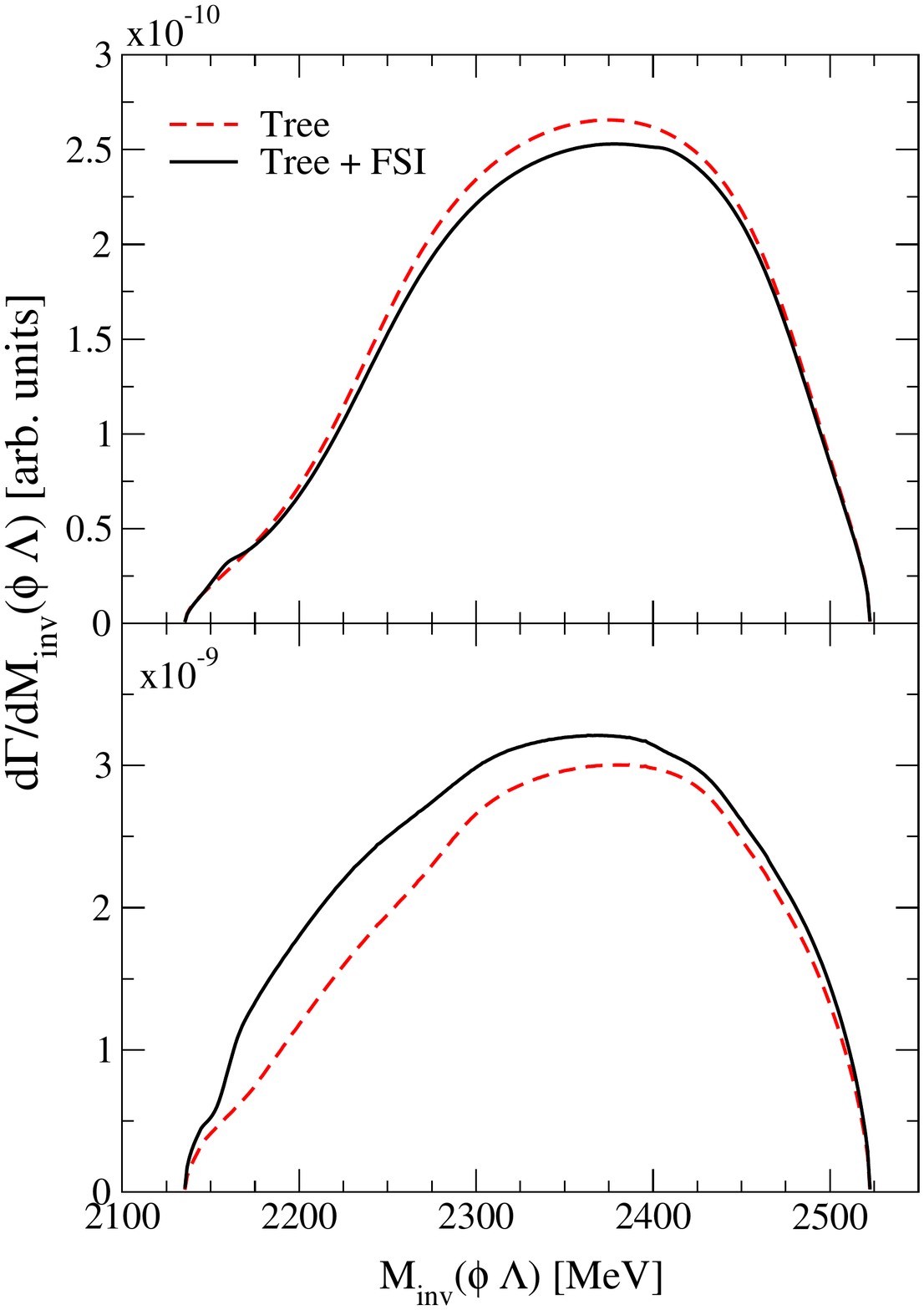}
    \caption{The $\phi\, \Lambda$ invariant mass spectrum computed with one broad $X(4140)$ resonance (top panel) or with a narrow $X(4140)$ plus a $X(4160)$ (bottom panel). The dashed red curve corresponds to considering only the tree level diagram while the solid black curve takes into account the final state interaction.}
    \label{phi_Lambda_spec}
\end{figure}

We finally present our results for the $\phi \, \Lambda$ invariant mass spectrum in Fig. \ref{phi_Lambda_spec}. Similar to our discussion of the $J/\psi \, \Lambda$ spectrum, we present calculations done with one broad $X(4140)$ resonance (top panel) jointly with those done in the presence of both the $X(4140)$ and $X(4160)$ (bottom panel). 

\begin{figure}[t]
    \centering
    \includegraphics[scale=0.36]{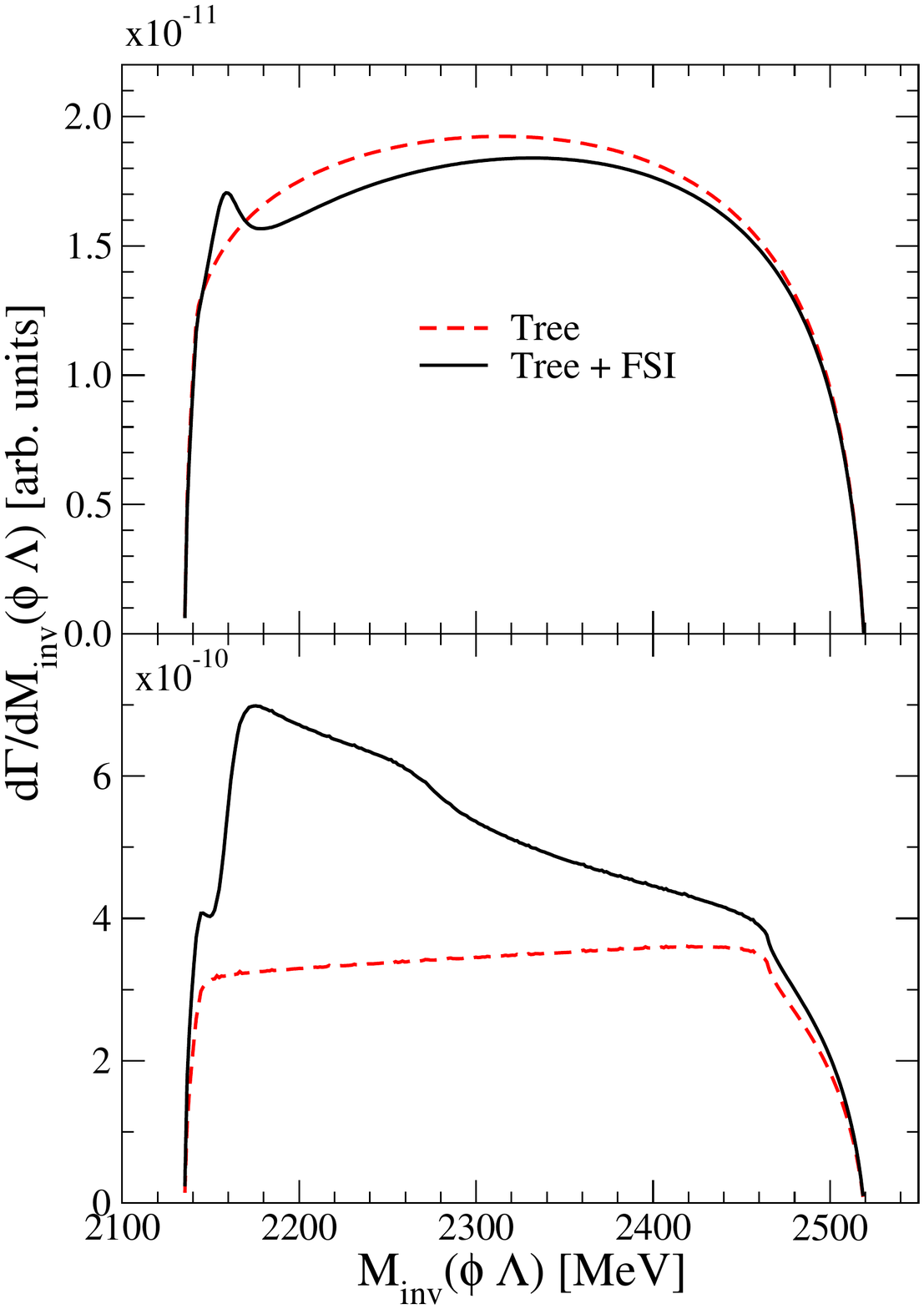}
    \caption{Same as Fig. \ref{phi_Lambda_spec} but with the constraint $M_{J/\psi \phi} > 4300$ MeV.}
    \label{backcut_phi_L}
\end{figure}

In Fig. \ref{phi_Lambda_spec} the red dashed curve displays the background obtained with only the tree level diagram $\mathcal{M}^{P}_{X_{4140}}$. The black solid curve is the result of the calculation that includes in addition the final state interaction terms  $\mathcal{M}^{\phi \Lambda}_{X_{4140}}$ and $\mathcal{M}^{J/\psi \Lambda}_{X_{4140}}$. Since the $\mathcal{M}^{\phi \Lambda}_{X_{4140}}$ term is responsible for the appearance of the resonance predicted in Ref. \cite{oset}, we expect to see a signal in the black solid curve around $2160$ MeV. However, 
 the resonance only appears as a very tiny bump with respect to the background regardless of whether we consider one or two $X$ resonances. The reason lies, essentially, in the small coupling of this resonance to the $\phi \, \Lambda$ channel, of value $0.5 + i0.3$ as reported in Ref.~\cite{oset}.

Since we do not have a strong presence of the $\phi \, \Lambda$ resonance in Fig. \ref{phi_Lambda_spec}, we now present an additional analysis for the $\phi \, \Lambda$ spectrum similar to the one presented for the $J/\psi \, \Lambda$ one, i.e., we impose certain constraints on the $J/\psi \, \phi$ invariant mass to reduce the background with respect to the expected peak. Fig. \ref{backcut_phi_L} displays the results of one such calculation with the constraint on the $J/\psi \, \phi$ invariant mass being $M_{J/\psi \phi} > 4300$ MeV to  avoid integrating over the dominant $X$ resonance(s). Although, as seen in Fig. \ref{backcut_phi_L}, the background is indeed drastically modified and the resonance becomes more visible, its signal is still not quite strong, leading us to the conclusion that there is little hope of seeing this resonance in this experiment. However this should not discourage the study of the $\phi \ \Lambda$ spectrum because, as can be seen in both Fig. \ref{phi_Lambda_spec} and Fig. \ref{backcut_phi_L}, there are differences in the obtained spectrum depending on whether we consider one or two $X$ resonances. This means further information on the $J/\psi \ \phi$ spectrum can be potentially obtained by the study of the $\phi \ \Lambda$ distribution. 

\section{Conclusion}

The recent discovery of various exotic hadrons \cite{Belle1,CDF1,CDF2,LHCb,CMS,D01,D02,PDG2,exp_X,exp_X_another,aaij,LHCnew} has made this topic a very exciting field of research. These discoveries, complemented by various theoretical studies \cite{41401,41402,41403,41404,X(4160),wang,Wu1,Wu2},  motivated us to study the decay $\Lambda_b \rightarrow J/\psi \ \phi \ \Lambda$, a reaction particularly well suited to identify several exotic hadrons as one can, in principle, hope for some interesting observation in every two-particle channel. 

The general motivation for studying the $\Lambda_b \rightarrow J/\psi \ \phi \ \Lambda$ decay comes from the Dalitz plots shown in Section 2. Depicting the available phase space in this reaction provided a clear idea of those resonances that could appear in the two-body mass spectra, and we identified several experimental and theoretical predicted candidates. Although the decay model we subsequently developed may not account for all of them, the Dalitz plots provided a good argument for experimentally measuring the $\Lambda_b \rightarrow J/\psi \ \phi \ \Lambda$ decay.

The chiral, coupled-channels approach complemented with various unitarization techniques has proved to be an extremely useful tool in the field of hadronic physics and is the one used repeatedly in this work. Equipped with this formalism, we have built a decay model for the $\Lambda_b \rightarrow J/\psi \ \phi \ \Lambda$ decay that is sufficiently rich to account for resonances, such as the $X(4140)$, the $X(4160)$, a $\Lambda (2160)$ resonance predicted in Ref. \cite{oset}, and a strange pentaquark predicted in Refs. \cite{Wu1,Wu2}. Our model is specifically designed to study the interplay between the $X(4140)$ and the $X(4160)$ resonances and also takes into account the final state interaction between $J/\psi \, \Lambda$ and $\phi \, \Lambda$ pairs.

In the $J/\psi \, \phi$ spectrum we have enunciated the differences between two approaches: one in which we have a single broad $X(4140)$ state and another in which we have two resonances, a narrow $X(4140)$ plus a wide $X(4160)$. In the latter case, the most interesting feature is the presence of a prominent cusp at the $D_s^* \, \Bar{D}_s^*$ threshold which strongly favors a molecular interpretation for the $X(4160)$. A similar cusp was obtained in the $B^+ \rightarrow J/\psi \ \phi \ K^+$ decay study of Ref.~\cite{wang}, from which we obtain clues for the value of the relative weight factor between the $X(4140)$ and the $X(4160)$. Quite encouragingly, the cusp persists despite variations in this relative weight and should thus provide an incentive for experimental collaborations to include such a threshold behaviour in the analysis of their data. In this work, we have focused only on the low energy regime of the $J/\psi \, \phi$ mass spectrum as our present model is not able to account for the other resonances that might appear at higher energies, such as $X(4274),X(4350)$ and $X(4500)$. In the future, it might be interesting to add these resonances to our model and thus provide a more quantitative analysis.

In the $J/\psi \ \Lambda$ spectrum, we have studied in detail the possibility of observing the strange pentaquark, predicted in a chiral, unitary, coupled-channels approach. In this work we have used a Breit-Wigner representation for the pentaquark, which allows us to keep the model simple and also study variations of its parameters such as its pole position and coupling to the $J/\psi \, \Lambda$ channel. The conclusion is that, if the pentaquark has a mass in the range $4450-4500$ MeV, it has a good chance to be experimentally detected in this decay mode but this may deteriorate quickly with increasing values of mass or with very low values of couplings.

We also studied the $\phi \, \Lambda$ spectrum, where a dynamically generated resonance was predicted. This resonance shows up quite weakly in the mass spectrum and has little chance of experimental detection.

We have also demonstrated that in both the $J/\psi \ \Lambda$ and $\phi \, \Lambda$ spectra, one obtains qualitative differences depending on whether we consider one or two $X$ resonances in the $J/\psi \ \phi$ mass distribution. This link between the different two-body invariant mass spectra is quite interesting as the study of one can illuminate the features of another.

In summary, our study has shown that the  $\Lambda_b \to
 J/\psi \ \phi \ \Lambda$ decay is a very promising reaction because it may give signatures of several exotic hadrons in their various two-body invariant mass spectra. We strongly encourage experimental collaborations to study this decay as it will open new windows to further learn about the dynamics of hadrons and their nature.

\section*{Acknowledgments}
This work is partly supported by the Spanish Ministerio de Economia y Competitividad (MINECO) under the project MDM-2014-0369 of ICCUB (Unidad de Excelencia 'Mar\'\i a de Maeztu'), 
and, with additional European FEDER funds, under the contract
FIS2017-87534-P.\\

\appendix
\section{Double Loop Integrals}
\label{DoubleLoopIntegrals}

When discussing the decay model for the reaction $\Lambda_b \rightarrow J/\psi \ \phi \ \Lambda$, we came across two final-state interaction amplitudes, $ \mathcal{M}^{J/\psi \Lambda}_{X_{4160}}$ and $ \mathcal{M}^{\phi \Lambda}_{X_{4160}}$, that involve evaluating a loop integral that is a 3-vector. The formal expression of the loop integral is quite similar in both cases and in this appendix we evaluate the former, briefly sketching at the end how to obtain the latter. Also, based on the discussion here, it should be obvious how to perform the integrals appearing in the expression $ \mathcal{M}^{J/\psi \Lambda}_{X_{4140}}$ and $ \mathcal{M}^{\phi \Lambda}_{X_{4140}}$, which are scalars.

\begin{figure}[h]
\centering
\includegraphics[scale=0.94]{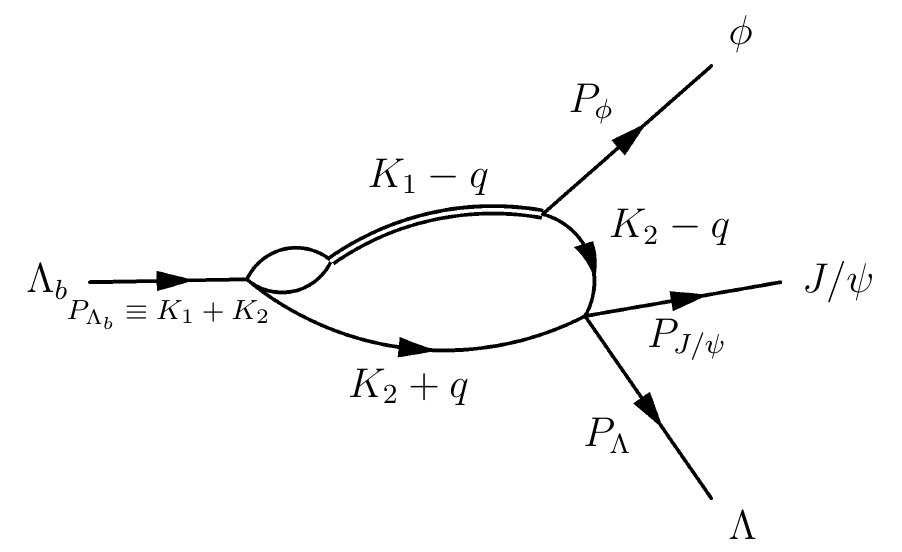}
    \caption{Feynman graph associated to the amplitude $ \mathcal{M}^{J/\psi \Lambda}_{X_{4160}}$.}
    \label{loop}
\end{figure}

Fig. \ref{loop} shows the diagram associated with the amplitude $ \mathcal{M}^{J/\psi \Lambda}_{X_{4160}}$. Here, $K_1= ({P_{\Lambda_B}+P_{\phi}})/{2}$ and $K_2= ({P_{\Lambda_B}-P_{\phi}})/{2}$. The labelling of the momenta in such a manner will make calculations easier later on. We start by writing down the expression for $ \mathcal{M}^{J/\psi \Lambda}_{X_{4160}}$ based on Fig. \ref{loop},

\begin{widetext}
\begin{equation}
\begin{split}
     \mathcal{M}^{J/\psi \Lambda}_{X_{4160}} = iT_{J/\psi \Lambda,J/\psi \Lambda} & \int \frac{d^4q}{(2 \pi)^4}  \frac{(\Vec{\epsilon}_{J/\psi}\times \Vec{\epsilon}_{\phi}) \cdot (\Vec{K}_2+\Vec{q}\,) \ G_{D_s^*\Bar{D}_s^*}}{(K_1-q)^2-M_X^2+iM_X\Gamma_X}   \frac{1}{(K_2-q)^2-M_{J/\psi}^2+i\epsilon} \frac{2M_{\Lambda}}{(K_2+q)^2-M_{\Lambda}^2+i\epsilon}.
\end{split}
\end{equation}
\end{widetext}
Here $T_{J/\psi \Lambda,J/\psi \Lambda}$ represents the interaction between $J/\psi$ and $\Lambda$ which is taken to be a Breit-Wigner corresponding to the pentaquark. As we take a $P-$wave weak decay vertex, we have the corresponding $P-$wave operator involving the polarization vectors of $J/\psi$ and $\phi$. The three propagators inside the loop are those of the $\Lambda$, $J/\psi$ and the resonance $X(4160)$ which, for the sake of brevity, is denoted as $X$. Preparing to do first the integral over $q^0$, we write the above expression as  
\begin{equation}
\begin{split}
    \mathcal{M}^{J/\psi \Lambda}_{X_{4160}} &= iT_{J/\psi \Lambda,J/\psi \Lambda} (\Vec{\epsilon}_{J/\psi}\times \Vec{\epsilon}_{\phi}) \cdot  \int  \frac{d^3q}{(2 \pi)^3} (\Vec{K}_2+\Vec{q}\,) \\ &\quad \int \frac{dq^0}{2 \pi} \frac{G_{D_s^*\Bar{D}_s^*}}{(K_1^0-q^0)^2-\omega_X^2+iM_X\Gamma_X} \\
    &\quad \frac{1}{(K_2^0-q^0)^2-\omega_{J/\psi}^2+i\epsilon} \frac{2M_{\Lambda}}{(K_2^0+q^0)^2-\omega_{\Lambda}^2+i\epsilon},
\end{split}
\end{equation}
where $\omega_X^2 = M_X^2 + (\Vec{K}_1-\Vec{q}\,)^2$, $\omega_{J/\psi}^2 = M_{J/\psi}^2 + (\Vec{K}_2-\Vec{q}\,)^2$ and $\omega_{\Lambda}^2 = M_{\Lambda}^2 + (\Vec{K}_2+\Vec{q}\,)^2$. Taking the non-relativistic limit,
\begin{equation}
\begin{split}
    \mathcal{M}^{J/\psi \Lambda}_{X_{4160}} & = iT_{J/\psi \Lambda,J/\psi \Lambda} (\Vec{\epsilon}_{J/\psi}\times \Vec{\epsilon}_{\phi}) \cdot  \int  \frac{d^3q}{(2 \pi)^3} (\Vec{K}_2+\Vec{q}\,) \\  &
    \quad 
    \int \frac{dq^0}{2 \pi} \frac{G_{D_s^*\Bar{D}_s^*}}{2\omega_X[K_1^0-q^0-\omega_X+i\frac{\Gamma_X}{2}]}   \\ 
    & \quad\quad\quad\quad \frac{1}{2\omega_{J/\psi}} \frac{1}{[K_2^0-q^0-\omega_{J/\psi}+i\epsilon]} \\
    & \quad\quad\quad\quad\ \frac{M_{\Lambda}}{\omega_{\Lambda}}\frac{1}{[K_2^0+q^0-\omega_{\Lambda}+i\epsilon]} \ .
\end{split}
\end{equation}
We will perform the integral over $q^0$ by the method of contour integration and therefore enumerate the poles for $q^0$: 
\begin{eqnarray}
        &q^0 = K_1^0 - \omega_X + i\frac{\Gamma_X}{2}  \ \ \ \ \ \ \ \ \text{Resonance},\\
        &\quad q^0 = K_2^0 - \omega_{J/\psi} + i\epsilon \ \ \ \ \ \text{$J/\psi$ propagator},\\
        &\quad q^0 = -K_2^0 + \omega_{\Lambda} - i\epsilon \ \ \ \ \ \text{$\Lambda$ propagator}.
\end{eqnarray}

In the complex plane, there are two poles above the real axis and one below. We choose to close the contour in the lower half plane thereby picking the one below the real axis ($\Lambda$ pole), finding
\begin{equation}
\begin{split}
    \mathcal{M}^{J/\psi \Lambda}_{X_{4160}} &= T_{J/\psi \Lambda,J/\psi \Lambda} (\Vec{\epsilon}_{J/\psi}\times \Vec{\epsilon}_{\phi}) \cdot  \\ &\quad \int  \frac{d^3q}{(2 \pi)^3}   \frac{(\Vec{K}_2+\Vec{q}\,) \ G_{D_s^*\Bar{D}_s^*}}{2\omega_X[K_1^0+K_2^0 - \omega_{\Lambda}-\omega_X+i\frac{\Gamma_X}{2}]} \\
    &\quad \frac{1}{2\omega_{J/\psi}[2K_2^0 - \omega_{\Lambda}-\omega_{J/\psi}+i\epsilon]} \frac{M_{\Lambda}}{\omega_{\Lambda}}.
\end{split}
\end{equation}

We now deal with the vector nature of the integral in the following way. In the Jackson frame (the $J/\psi \ \phi$ rest frame) the vectors $\Vec{K}_1$ and $\Vec{K}_2$ are approximately equal to each other as $\Vec{P}_{\phi}$ is small. Further  $\Vec{K}_1$ appears only in the expression $\omega_X^2 = M_X^2 + (\Vec{K}_1-\Vec{q}\,)^2$ and therefore we expect that the effects of $\Vec{K}_1$ will be supressed by the large value of $M_X$. Thus, as an approximation, we replace $\Vec{K}_1$ with $\Vec{K}_2$. Now, $\omega_X^2 = M_X^2 + (\Vec{K}_2-\Vec{q}\,)^2$ and there is only one 3-vector appearing in the integrand ($\Vec{K}_2$) and therefore we take the integral to be proportional to this vector. This gives,
\begin{equation}
    \mathcal{M}^{J/\psi \Lambda}_{X_{4160}} = T_{J/\psi \Lambda,J/\psi \Lambda} (\Vec{\epsilon}_{J/\psi}\times \Vec{\epsilon}_{\phi}) \cdot \Vec{K}_2 \  I^{J/\psi \Lambda}_{X_{4160}},
\end{equation}
with,
\begin{eqnarray}
        I^{J/\psi \Lambda}_{X_{4160}} &&=  \int  \frac{d^3q}{(2 \pi)^3}   \frac{\Vec{K}_2 \cdot (\Vec{K}_2+\Vec{q}\,) \ G_{D_s^*\Bar{D}_s^*}}{2\omega_X[K_1^0+K_2^0 - \omega_{\Lambda}-\omega_X+i\frac{\Gamma_X}{2}]} \nonumber \\
   & & \frac{1}{|\Vec{K}_2|^2} \ \  \frac{1}{2\omega_{J/\psi}[2K_2^0 - \omega_{\Lambda}-\omega_{J/\psi}+i\epsilon]} \frac{M_{\Lambda}}{\omega_{\Lambda}} .
\end{eqnarray}
To make things simpler we make the shift $\Vec{q} \rightarrow \Vec{K}_2+\Vec{q}$ which gives,
\begin{eqnarray}
        I^{J/\psi \Lambda}_{X_{4160}} &&=  \int  \frac{d^3q}{(2 \pi)^3}   \frac{\Vec{K}_2 \cdot \Vec{q} \ \  G_{D_s^*\Bar{D}_s^*}}{2\omega_X[K_1^0+K_2^0 - \omega_{\Lambda}-\omega_X+i\frac{\Gamma_X}{2}]} \nonumber \\
    &&\frac{1}{|\Vec{K}_2|^2} \ \ \frac{1}{2\omega_{J/\psi}[2K_2^0 - \omega_{\Lambda}-\omega_{J/\psi}+i\epsilon]} \frac{M_{\Lambda}}{\omega_{\Lambda}},
    \label{b8}
\end{eqnarray}
where $\omega_X^2 = M_X^2 + (2\Vec{K}_2-\Vec{q}\,)^2$, $\omega_{J/\psi}^2 = M_{J/\psi}^2 + (2\Vec{K}_2-\Vec{q}\,)^2$ and $\omega_{\Lambda}^2 = M_{\Lambda}^2 + \Vec{q}\,^2$. 

Having completed the evaluation of $\mathcal{M}^{J/\psi \Lambda}_{X_{4160}}$, it is obvious how to obtain $\mathcal{M}^{\phi \Lambda}_{X_{4160}}$ as we only need to replace $J/\psi$ with $\phi$ and vice versa. In the Jackson frame, this means replacing $\Vec{K}_2$ with $\Vec{K}_1$ and we obtain
\begin{eqnarray}
    \mathcal{M}^{\phi \Lambda}_{X_{4160}} = T_{\phi \Lambda,\phi \Lambda} (\Vec{\epsilon}_{J/\psi}\times \Vec{\epsilon}_{\phi}) \cdot \Vec{K}_1 \  I^{\phi \Lambda}_{X_{4160}},
\end{eqnarray}
with
\begin{eqnarray}
        I^{\phi \Lambda}_{X_{4160}} &&=   \int  \frac{d^3q}{(2 \pi)^3}   \frac{\Vec{K}_1 \cdot \Vec{q} \ \  G_{D_s^*\Bar{D}_s^*}}{2\omega_X[K_1^0+K_2^0 - \omega_{\Lambda}-\omega_X+i\frac{\Gamma_X}{2}]} \nonumber \\
    && \frac{1}{|\Vec{K}_1|^2} \ \ \frac{1}{2\omega_{\phi}[2K_1^0 - \omega_{\Lambda}-\omega_{\phi}+i\epsilon]} \frac{M_{\Lambda}}{\omega_{\Lambda}},
    \label{b10}
\end{eqnarray}
where $\omega_X^2 = M_X^2 + (2\Vec{K}_1-\Vec{q}\,)^2$, $\omega_{\phi}^2 = M_{\phi}^2 + (2\Vec{K}_1-\Vec{q}\,)^2$ and $\omega_{\Lambda}^2 = M_{\Lambda}^2 + \Vec{q}\,^2$. 

Finally, we also give the expressions for the scalar loop integrals appearing in $\mathcal{M}^{J/\psi \Lambda}_{X_{4140}}$ and $ \mathcal{M}^{\phi \Lambda}_{X_{4140}}$, namely 
\begin{eqnarray}
        I^{J/\psi \Lambda}_{X_{4140}} &&=  \int  \frac{d^3q}{(2 \pi)^3}   \frac{1}{2\omega_X[K_1^0+K_2^0 - \omega_{\Lambda}-\omega_X+i\frac{\Gamma_X}{2}]} \nonumber \\
    && \frac{1}{2\omega_{J/\psi}[2K_2^0 - \omega_{\Lambda}-\omega_{J/\psi}+i\epsilon]} \frac{M_{\Lambda}}{\omega_{\Lambda}},
    \label{b11}
\end{eqnarray}
where $\omega_X^2 = M_X^2 + (2\Vec{K}_2-\Vec{q}\,)^2$, $\omega_{J/\psi}^2 = M_{J/\psi}^2 + (2\Vec{K}_2-\Vec{q}\,)^2$ and $\omega_{\Lambda}^2 = M_{\Lambda}^2 + \Vec{q}\,^2$, and
\begin{eqnarray}
        I^{\phi \Lambda}_{X_{4140}}&& =  \int  \frac{d^3q}{(2 \pi)^3}   \frac{1}{2\omega_X[K_1^0+K_2^0 - \omega_{\Lambda}-\omega_X+i\frac{\Gamma_X}{2}]} \nonumber \\
    && \frac{1}{2\omega_{\phi}[2K_1^0 - \omega_{\Lambda}-\omega_{\phi}+i\epsilon]} \frac{M_{\Lambda}}{\omega_{\Lambda}},
    \label{b12}
\end{eqnarray}
where $\omega_X^2 = M_X^2 + (2\Vec{K}_1-\Vec{q})^2$, $\omega_{\phi}^2 = M_{\phi}^2 + (2\Vec{K}_1-\Vec{q})^2$ and $\omega_{\Lambda}^2 = M_{\Lambda}^2 + \Vec{q}^2$.

The derivation of Eqs. (\ref{b11}) and (\ref{b12}) is entirely analogous to that of Eqs. (\ref{b8}) and (\ref{b10}) with the only differences being the absence of the $D_s^* \ \Bar{D}_s^*$ loop function and a few 3-momentum factors.

\section{Spin Sums}
\label{SpinSums}

In our discussion of the decay model for the reaction $\Lambda_b \rightarrow J/\psi \ \phi \ \Lambda$, we had collected all the amplitudes involving the $X(4160)$ as follows
\begin{eqnarray}
    \mathcal{M}_{X_{4160}} &=&(\Vec{\epsilon}_{J/\psi}\times \Vec{\epsilon}_{\phi}) \cdot  \big( \Vec{P}_{\Lambda} \mathcal{\Tilde{M}}^{P}_{X_{4160}}
    + \Vec{K}_2   \mathcal{\Tilde{M}}^{J/\psi \Lambda}_{X_{4160}} \nonumber \\ && +
    \Vec{K}_1 \mathcal{\Tilde{M}}^{\phi \Lambda}_{X_{4160}} \big),
\end{eqnarray}
where we had defined $\Vec{K}_2 \equiv (\Vec{P}_{\Lambda}-\Vec{P}_{\phi})/2$ and $\Vec{K}_1 \equiv (\Vec{P}_{\Lambda}+\Vec{P}_{\phi})/2$.

In this appendix we demonstrate how to perform the sum over the polarizations $\Vec{\epsilon}_{J/\psi}$ and $\Vec{\epsilon}_{\phi}$. First we need to take the square of the absolute value of $\mathcal{M}_{X_{4160}}$ which is
\begin{eqnarray}
        |\mathcal{M}_{X_{4160}}|^2 &=& \varepsilon^{ijk}\, \epsilon^{j}_{J/\psi} \, \epsilon^{k}_{\phi} \ \ \varepsilon^{abc} \, \epsilon^{b}_{J/\psi} \, \epsilon^{c}_{\phi}  \nonumber \\
         &&
         \big(P^{i}_{\Lambda} \mathcal{\Tilde{M}}^{P}_{X_{4160}}
    + K^{i}_2   \mathcal{\Tilde{M}}^{J/\psi \Lambda}_{X_{4160}} +
    K^{i}_1 \mathcal{\Tilde{M}}^{\phi \Lambda}_{X_{4160}} \big)  \nonumber \\
    &&
        \big(P^{a}_{\Lambda} \mathcal{\Tilde{M}}^{* P}_{X_{4160}}
    + K^{a}_2   \mathcal{\Tilde{M}}^{* J/\psi \Lambda}_{X_{4160}} +
    K^{a}_1 \mathcal{\Tilde{M}}^{* \phi \Lambda}_{X_{4160}} \big) \ . \nonumber \\
    \label{c2}
\end{eqnarray}
Note that we have switched to an index based notation under which repeated indices are summed over. The indices $i$, $j$, $k$, $a$, $b$ and $c$, which appear as superscripts, label the components of the various Cartesian tensors. Also, $\varepsilon^{ijk}$ is the Levi-Civita symbol and should be distinguished from the polarization vectors. 

Recall that we are working in the frame in which the $J/\psi \ \phi$ system is at rest. Here, the 3-momenta of $J/\psi$ and $\phi$ are small relative to their masses and thus we can take the non-relativistic limit. The sum over the $J/\psi$ and $\phi$ polarizations are then expressed as
\begin{eqnarray}
    &&\sum_{\text{pol}} \epsilon^{j}_{J/\psi} \ \epsilon^{b}_{J/\psi} = \delta^{jb},
    \label{c3} \\
    &&\sum_{\text{pol}} \epsilon^{k}_{\phi} \ \epsilon^{c}_{\phi} = \delta^{kc}.
    \label{c4}
\end{eqnarray}
Thus when we perform the sum over the polarizations in Eq. (\ref{c2}) we get,
\begin{eqnarray}
        \overline{|\mathcal{M}_{4160}|^2} &= \varepsilon^{ijk}  \  
        \big(P^{i}_{\Lambda} \mathcal{\Tilde{M}}^{P}_{X_{4160}}
    + K^{i}_2   \mathcal{\Tilde{M}}^{J/\psi \Lambda}_{X_{4160}} +
    K^{i}_1 \mathcal{\Tilde{M}}^{\phi \Lambda}_{X_{4160}} \big)  \nonumber \\
    &\quad 
        \varepsilon^{ajk}  \big(P^{a}_{\Lambda} \mathcal{\Tilde{M}}^{* P}_{X_{4160}}
    + K^{a}_2   \mathcal{\Tilde{M}}^{* J/\psi \Lambda}_{X_{4160}} +
    K^{a}_1 \mathcal{\Tilde{M}}^{* \phi \Lambda}_{X_{4160}} \big) \ .  \nonumber \\
    \label{c5}
\end{eqnarray}
Using the identity $\varepsilon^{ijk} \ \varepsilon^{ajk} = \delta^{ia}$, to write 
\begin{eqnarray}
        \overline{|\mathcal{M}_{4160}|^2} &= \delta^{ia} \  
        \big(P^{i}_{\Lambda} \mathcal{\Tilde{M}}^{P}_{X_{4160}}
    + K^{i}_2   \mathcal{\Tilde{M}}^{J/\psi \Lambda}_{X_{4160}} +
    K^{i}_1 \mathcal{\Tilde{M}}^{\phi \Lambda}_{X_{4160}} \big) \nonumber \\
    &\quad 
        \big(P^{a}_{\Lambda} \mathcal{\Tilde{M}}^{* P}_{X_{4160}}
    + K^{a}_2   \mathcal{\Tilde{M}}^{* J/\psi \Lambda}_{X_{4160}} +
    K^{a}_1 \mathcal{\Tilde{M}}^{* \phi \Lambda}_{X_{4160}} \big) \ ,  \nonumber \\
    \label{c6}
\end{eqnarray}
which allows us to go back to the original notation without indices
\begin{eqnarray}
        \overline{|\mathcal{M}_{4160}|^2} &= \big( \Vec{P}_{\Lambda} \mathcal{\Tilde{M}}^{P}_{X_{4160}}
    + \Vec{K}_2   \mathcal{\Tilde{M}}^{J/\psi \Lambda}_{X_{4160}} +
    \Vec{K}_1 \mathcal{\Tilde{M}}^{\phi \Lambda}_{X_{4160}} \big) \nonumber \\
    &\quad \boldsymbol{\cdot} \big( \Vec{P}_{\Lambda} \mathcal{\Tilde{M}}^{* P}_{X_{4160}}
    + \Vec{K}_2   \mathcal{\Tilde{M}}^{* J/\psi \Lambda}_{X_{4160}} +
    \Vec{K}_1 \mathcal{\Tilde{M}}^{* \phi \Lambda}_{X_{4160}} \big) \ . \nonumber \\
\end{eqnarray}
Carrying out the dot product gives the expression of Eq.~(\ref{5.17}) in the text.

\end{document}